\documentclass[11pt,a4paper]{article}
\usepackage{jcappub}
\usepackage{amsmath}
\usepackage{epsfig}
\usepackage{bm}
\usepackage{wasysym}
\usepackage{mathtools}
\usepackage{footnote}
\usepackage{relsize}
\usepackage{graphicx}
\usepackage{epsfig}
\usepackage{epstopdf}
\usepackage[font=normalsize,justification=raggedright,format=hang]{caption}
\usepackage{subcaption}
\usepackage{hyperref}

\newcommand*{\GtrSim}{\smallrel\gtrsim}
\newcommand*{\LessSim}{\smallrel\lesssim}

\makeatletter
\newcommand*{\smallrel}[2][.8]{%
  \mathrel{\mathpalette{\smallrel@{#1}}{#2}}%
}
\newcommand*{\smallrel@}[3]{%
  \sbox0{$#2\vcenter{}$}%
  \dimen@=\ht0 %
  \raise\dimen@\hbox{%
    \scalebox{#1}{%
      \raise-\dimen@\hbox{$#2#3\m@th$}%
    }%
  }%
}
\makeatother

\newcommand*\diff{\mathop{}\!\mathrm{d}}
\newcommand*\Diff[1]{\mathop{}\!\mathrm{d^#1}}
\newcommand{\B}[1]{\mathbf{#1}}
\newcommand{\T}[1]{\mathrm{#1}}
\DeclareMathOperator{\erf}{erf}
\DeclareMathOperator{\e}{e}
\makeatletter
\newcommand{\pushright}[1]{\ifmeasuring@#1\else\omit\hfill$\displaystyle#1$\fi\ignorespaces}
\newcommand{\pushleft}[1]{\ifmeasuring@#1\else\omit$\displaystyle#1$\hfill\fi\ignorespaces}
\makeatother

\begin{document}

\title{Monthly Modulation in Dark Matter Direct-Detection Experiments}
\author[a,b]{Vivian Britto}
\author[a]{and Joel Meyers}
\affiliation[a]{Canadian Institute for Theoretical Astrophysics, University of Toronto, Toronto, Ontario M5S 3H8, Canada}
\affiliation[b]{Department of Physics, University of Toronto, Toronto, Ontario M5S 1A7, Canada}
\emailAdd{vivian.britto@mail.utoronto.ca}
\emailAdd{jmeyers@cita.utoronto.ca}

\date{\today}

\abstract{
The signals in dark matter direct-detection experiments should exhibit modulation signatures due to the Earth's motion with respect to the Galactic dark matter halo. The annual and daily modulations, due to the Earth's revolution about the Sun and rotation about its own axis, have been explored previously. Monthly modulation is another such feature present in direct detection signals, and provides a nearly model-independent method of distinguishing dark matter signal events from background. We study here monthly modulations in detail for both WIMP and WISP dark matter searches, examining both the effect of the motion of the Earth about the Earth-Moon barycenter and the gravitational focusing due to the Moon. For WIMP searches, we calculate the monthly modulation of the count rate and show the effects are too small to be observed in the foreseeable future.  For WISP dark matter experiments, we show that the photons generated by WISP to photon conversion have frequencies which undergo a monthly modulating shift which is detectable with current technology and which cannot in general be neglected in high resolution WISP searches.
}

\arxivnumber{1409.2858}

\maketitle
\section{Introduction} \label{introduction}
There exists a preponderance of evidence that some form of non-baryonic dark matter makes up a significant fraction of the mass in the universe; its detailed properties however, remain a mystery \cite{Agashe:2014kda}.  Understanding the nature of dark matter holds considerable importance for particle physics, high energy theory, astrophysics, and cosmology, and there has hence been a great deal of effort expended toward achieving a better understanding of dark matter through experiment. Several intriguing hints notwithstanding, there so far have been no conclusive results to this end.

Many existing models for particle dark matter amenable to direct searches can be broadly divided into two classes: Weakly Interacting Massive Particles (WIMPs) with typical masses in the GeV to TeV range \cite{Jungman:1995df,Bertone:2004pz}, and Weakly Interacting Sub-eV Particles (WISPs) with masses below an eV \cite{Arias:2012az,Ringwald:2012hr,Essig:2013lka}.  Candidates for WIMPs include the lightest supersymmetric particle and the lightest Kaluza-Klein particle, while WISP candidates include axions, axion-like particles, and hidden photons.  Despite the different underlying physics and production mechanisms, WIMPs and WISPs are both expected to be cold and nearly collisionless today, making both classes of particles viable candidates for the cold dark matter that is present in our universe.  Furthermore, the dynamics of structure formation are largely insensitive to the detailed nature of cold dark matter, and so we can predict the properties of dark matter halos without restricting ourselves to a particular class of models.

Dark matter direct-detection experiments seek to measure the interaction of dark matter particles streaming through the Earth with standard model particles present in a detector.  Such experiments are clearly sensitive to the differing properties of dark matter, and so we must distinguish experiments aimed at detecting WIMPs from those which attempt to detect WISPs.  For WIMP dark matter searches, collisions of dark matter particles with the atoms in the detector material may result in ionization along with the deposit of heat and/or light, which are registered by the detector \cite{Goodman:1984dc,Smith:1988kw,Lewin:1995rx}.  Many such experiments are currently in operation around the world (see Sec.\ 25 of \cite{Agashe:2014kda} for a recent survey of the experimental situation).  On the other hand, experiments which look for WISP dark matter attempt to detect their conversion to photons in an electromagnetic cavity permeated by a large static magnetic field\footnote{The magnetic field is irrelevant for hidden photons but is a necessary component for axion-like particles.}.  A small but increasing number of WISP dark matter detection experiments are currently underway \cite{Asztalos:2001tf,Asztalos:2009yp,Slocum:2014gwa,Horns:2014qta}.

One of the main challenges that WIMP detection experiments face is the presence of backgrounds. For instance, cosmic rays and radioactive decays in the material in and around the experiment can create signals which mimic those of dark matter scattering events.  Due to such effects, experiments are often carried out underground with large amounts of shielding, and also employ sophisticated methods for distinguishing background events from candidate signal events. Despite these efforts to minimize the number of background events, they can never be completely removed from the data.

One method which has been proposed to distinguish dark matter scattering events from the background is to study the time dependence of the event rate \cite{Drukier:1986tm,Freese:1987wu,Freese:2012xd}. Specifically, since the motion of a detector relative to the dark matter halo affects the observed event rate by altering the incoming flux of dark matter particles, there should be modulation signatures in the rate count whose details are largely independent of dark matter properties and detector physics, though they are sensitive to the local dark matter phase space distribution \cite{Drukier:1986tm,Freese:1987wu,Freese:2012xd}. Annual modulation is the most prominent example of such an effect, and it arises due to the annual motion of Earthbound detectors around the Sun.

The DAMA experiment has operated for more than a decade and has reported with high significance an annually modulating rate of dark matter candidate events \cite{Bernabei:2013xsa}.  Although the amplitude and phase of the modulation seem to agree well with the modulation expected for dark matter events in the simplest astrophysical scenarios, an interpretation in terms of dark matter seems to be at odds with the null results from several other direct-detection experiments \cite{Akerib:2013tjd,Aprile:2012nq,Angle:2011th,Agnese:2013rvf,Agnese:2013jaa,Armengaud:2011cy}. While it was possible for a time to evade the constraints with particular parameter choices in some specific WIMP models \cite{Savage:2008er}, the increased sensitivity and exposure of WIMP detectors has made such explanations of the signal much more difficult.  This disagreement has led some to propose that the annual modulation seen in DAMA data is due to background events which themselves modulate annually (see e.g. \cite{Ralston:2010bd,Blum:2011jf,Davis:2014cja}), though some of these proposals have been disputed by the DAMA collaboration and others \cite{Bernabei:2012wp,Bernabei:2014tqa,Barbeau:2014mla}. Several of these criticisms are grounded on the fact that many physical quantities (temperature, cosmic ray activity, solar neutrino flux) vary on an annual cycle, which could in principle result in an annually modulating event rate.  Addressing the veracity of these claims and counter-claims is outside the scope of this paper.

In addition to annual modulation, the daily rotation of the Earth results in a diurnal modulation of the dark matter flux \cite{Collar:1992qc,Ling:2004aj,Lee:2013xxa,Kouvaris:2014lpa}.  While a detection of a diurnal modulation accompanying an annual one would provide great support to the dark matter interpretation of any proposed detection, there are several quantities (similar to those detailed above) which change on a daily cycle, and could also conceivably cause a diurnally modulating background event rate.  Furthermore, the diurnal modulation of the dark matter event rate results from a combination of the diurnal cycle of the detector velocity, the gravitational focusing of dark matter due to the Earth, and eclipsing of the stream of dark matter particles which pass through the bulk of the Earth, making predictions for the specific form of the diurnal modulation more complicated and model-dependent than the annual modulation.

The Earth also undergoes a monthly motion due to its interaction with the Moon, and this ``wobble'' of the Earth about the Earth-Moon barycenter results in a monthly modulation of the dark matter event rate. As we shall see, the expected modulation, though unambiguously present and nearly model-independent, is quite small. In contrast to the daily and annual modulations however, far fewer potential sources of background change over a monthly cycle, making a detection of monthly modulation an extremely convincing confirmation of the detection of dark matter.\footnote{It is well known that tides undergo a monthly cycle, and though this seems the most readily identifiable process which could conceivably result in a monthly modulating background, it is not obvious how tides would affect the event rates of dark matter detectors.  The sun has a (latitude dependent) synodic rotation period of around 27 days \cite{1998ApJ...505..390S}, and could in principle affect detectors on Earth in various ways.  We thank Peter Sturrock for drawing our attention to this latter point.}  It is worth mentioning at the outset that even though the exposure required to detect monthly modulation far exceeds current experimental capabilities, fully characterizing the expected dark matter signal is useful even without a detection.

WISP detection experiments are also affected by the periodic motion of the Earth.  In particular, WISPs are converted into photons inside the detector, and the frequency of the photons is determined by the energy of the incoming WISP particles.  The motion of the Earth relative to the dark matter halo produces a non-trivial shift in the energy of the dark matter particles, and thus causes periodic shifts in the frequency of observed photons.  For experiments such as the Axion Dark Matter eXperiment (ADMX) \cite{Asztalos:2001tf,Asztalos:2009yp} which rely on resonant production of photons \cite{Sikivie:1983ip}, these shifts in frequency due to the annual and diurnal motion of the Earth must be accounted for in the search strategy \cite{Ling:2004aj}.  Broadband WISP searches which do not rely on resonant enhancement are less sensitive to these periodic frequency shifts, but are capable of directional detection of WISPs and can in principle observe periodic changes in the direction of the WISP flux \cite{Horns:2012jf}.  The monthly motion of the Earth also leads to a modulation of the frequency of the photons produced in WISP dark matter experiments.  Though smaller than the annual and diurnal frequency shifts, the monthly modulation is at a level which is detectable with present technology, and in general cannot be neglected in the analysis of the ADMX data.

Similarly to the WIMP case, the time dependence of the signal in WISP searches can be used to discriminate signal from background since terrestrial sources which could be confused for WISPs in the detector are not expected to share the same modulating frequency.  If WISP dark matter is detected, it will be very useful to study in detail the local velocity distribution, and a proper mapping of photon frequency onto dark matter velocity requires a detailed accounting of the motion of the Earth.  It is therefore important to understand what level of modulation is expected in WISP direct-detection experiments, including that due to the monthly motion of the Earth.

In this work, we study in detail the monthly modulation anticipated in dark matter direct-detection experiments. Sec.~\ref{notation} reviews the relevant background material and defines the notation used throughout the paper.  In Sec.~\ref{monthly}, we examine the monthly modulation expected in WIMP and WISP dark matter searches.   We examine two sources of monthly modulation in the signals:\ the motion of the Earth about the Earth-Moon barycenter, and gravitational focusing due to the Moon.  For the WIMP case, we demonstrate that the monthly modulation depends on both the sidereal and synodic monthly periods, leading to an annually-varying amplitude for the monthly modulation. Further, we show that the gravitational focusing due to the moon has a negligible impact on the monthly modulation signal.  For WISP searches, we show that the monthly motion of the Earth and the eccentricity of the Moon's orbit lead to periodic shifts in signal frequency which in general must be taken into account.  We conclude in Sec.~\ref{discussion}.  The details of the coordinates and velocities we use are specified in Appendix~\ref{coordinates}.

\section{Background and Notation}\label{notation}

Although they differ in the details, both WIMP and WISP dark matter direct-detection experiments are sensitive to the local density and velocity distribution of dark matter.  The expected signals in these experiments contain several sources of time dependence. The dark matter density and distribution function in the solar neighborhood, for instance, are in principle inherently time dependent, though we will assume that any variation in these quantities occurs on time scales much longer than the relevant observations, and can thus be neglected. A more important source of time dependence is the motion of the detector relative to the rest frame of the dark matter halo, which changes the portion of the distribution function which is sampled by the experiment. Hence, the signal depends on the velocity of the detector. As stated in Sec.~\ref{introduction}, the most prominent of this sort of effect is an annual modulation due to the Earth's motion around the Sun, though there also exists a daily modulation due to the rotation of the Earth and a monthly modulation due to the motion of the Earth about the Earth-Moon barycenter.  Also, the gravitational influence of bodies near the detector distorts the local dark matter density and velocity distribution, which gives the signal a dependence upon the position of the detector relative to these bodies.

Let us first examine the effect of the Earth's motion. Suppose the velocity distribution of the dark matter halo is given by some $\tilde{f}(\mathbf{v})$. If we ignore gravitational focusing, this distribution is related to the dark matter distribution in the Lab frame $f(\mathbf{v},t)$ by the Galilean transformation
\begin{equation}
	f(\mathbf{v},t)=\tilde{f}(\mathbf{v}+\mathbf{v}_{\mathrm{obs}}(t)) \, ,
	\label{boostedf}
\end{equation}
where $\mathbf{v}_{\mathrm{obs}}(t)$ is the velocity of the detector relative to the dark matter halo frame.  Setting aside the effect of the Earth's rotation, $\mathbf{v}_{\mathrm{obs}}(t)$ is given by
\begin{equation}
	\mathbf{v}_{\mathrm{obs}}(t)=\mathbf{v}_{\odot}+\mathbf{v}_{\textsc{es}}(t) \, ,
	\label{vobs}
\end{equation}
where $\mathbf{v}_{\odot} \approx (11,232,7)$~km/s  is the velocity of the Sun in Galactic coordinates \cite{Kerr:1986hz,Schoenrich:2009bx}, and $\mathbf{v}_{\textsc{es}}(t)$ is the velocity of the Earth in the Solar frame.\footnote{Throughout this work we use the notation $\mathbf{r}_\textsc{xy}$ to denote the position of \textsc{X} as measured from \textsc{Y}, and $\mathbf{v}_\textsc{xy}$ to denote the velocity of \textsc{X} relative to \textsc{Y}.  The letter \textsc{S} refers to the Sun, \textsc{E} to the Earth, \textsc{M} to the Moon, and \textsc{B} to the Earth-Moon barycenter.} This latter velocity vector undergoes periodic changes due to the motion of the Earth which results in periodic modulation of the signals expected in dark matter direct-detection experiments.

The most significant periodic motion of the Earth is the annual orbit around the Sun.  In addition, the Earth also undergoes a monthly motion due to the gravitational interaction with the Moon.  The barycenter for the Earth-Moon system is located on the line joining their centers at a distance of
\begin{equation}
\frac{r_{\textsc{em}}}{1+\frac{M_\textsc{e}}{M_\textsc{m}}} \approx 4661\ \mathrm{km}
\label{barydistance}
\end{equation}
(roughly three-fourths of the Earth's radius) from the center of the Earth. Here $r_{\textsc{em}}$ is the distance between the centers of the Earth and Moon, and $M_\textsc{e}$ and $M_\textsc{m}$ are their respective masses.\footnote{The precise description of the orbit of the Earth-Moon system is complicated by the non-negligible effect of the Sun's gravitational force on the system, but it will be sufficient for our purposes to treat the orbit of the Earth around the barycenter as an ellipse which is slightly inclined (by about $5.2^\circ $) relative to the ecliptic, with the orbital period of a sidereal month, $T_{\mathrm{sid}}\approx 27.32$~days. See Appendix~\ref{coordinates} for a more detailed description of the orbits.} Including the monthly motion of the Earth around the Earth-Moon barycenter $\mathbf{v}_{\textsc{eb}}$ in the velocity of the detector $\mathbf{v}_{\mathrm{obs}}$ above results in a monthly modulation in the detection signal in addition the annual one.

To introduce the effect of gravitational focusing, we use the fact that Liouville's theorem guarantees the constancy of the phase-space density of dark matter particles along their trajectories \cite{Lee:2013wza}. For dark matter particles passing near the Sun and arriving at the Earth, we therefore have
\begin{equation}
	\rho_{\chi} f(\mathbf{v},t)=\rho_{\infty}\tilde{f}(\mathbf{v}_{\odot}+\mathbf{v}_{\infty,\textsc{s}}[\mathbf{v}_{\textsc{es}}(t)+\mathbf{v}]) \ ,
\label{lville}
\end{equation}
where $\rho_{\infty}$ is the dark matter density asymptotically far away from Sun's gravitational well.  The function $\mathbf{v}_{\infty,\textsc{s}}[\mathbf{v}]$ is derived from the conservation of the Laplace-Runge-Lenz vector \cite{Alenazi:2006wu,Sikivie:2002bj} and specifies the velocity $\mathbf{v}_{\infty,\textsc{s}}$ a particle must have far away from the Sun in order to arrive at the Earth with a velocity $\mathbf{v}$ as measured in the Solar frame. As such, it describes the gravitational focusing effect of the Sun and is given by
\begin{equation}
	\mathbf{v}_{\infty,\textsc{s}}[\mathbf{v}]=\frac{v_{\infty,\textsc{s}}^2\,\mathbf{v} + v_{\infty,\textsc{s}}\,u_{\mathrm{esc},\textsc{s}}^2\,\mathbf{\hat{r}}_{\textsc{es}}/2 - v_{\infty,\textsc{s}}\,\mathbf{v} \left(\mathbf{v}\cdot\mathbf{\hat{r}}_{\textsc{es}}\right)}{v_{\infty,\textsc{s}}^2 + u_{\mathrm{esc},\textsc{s}}^2 /2 - v_{\infty,\textsc{s}} \left(\mathbf{v}\cdot\mathbf{\hat{r}}_{\textsc{es}}\right)} \ ,
\label{vinfs}
\end{equation}
where $v_{\infty,\textsc{s}}^2 = v^2 -  u_{\mathrm{esc},\textsc{s}}^2$ from conservation of energy, and  $u_{\mathrm{esc},\textsc{s}} = \sqrt{2GM_{\odot}/r_{\textsc{es}}(t)} \approx 40$~km/s is the escape velocity from the Sun near the Earth's orbit. Hence, Eq.\ (\ref{lville}) captures the complete picture of annual modulation in the velocity distribution of dark matter in the Lab frame; it includes the both the annually-modulating velocity of the Earth around the Sun, and the effect of the Sun's gravitational focusing.

Finally, just as in the case of the Sun, the gravitational well of the Moon distorts the local distribution of dark matter.  As a result, the position of the Moon relative to the Earth affects the scattering rate, and introduces an additional source of monthly modulation.  In the absence of other masses, the velocity $\mathbf{v}_{\infty,\textsc{m}}$ that a particle must have infinitely far from the Moon in order to have a velocity $\mathbf{v}$ (in the Lunar frame) at the position of the Earth, is given by straightforward modifications to Eq.~(\ref{vinfs}):
\begin{equation}
	\mathbf{v}_{\infty,\textsc{m}}[\mathbf{v}]=\frac{v_{\infty,\textsc{m}}^2\,\mathbf{v} + v_{\infty,\textsc{m}}\,u_{\mathrm{esc},\textsc{m}}^2\,\mathbf{\hat{r}}_{\textsc{em}}/2 - v_{\infty,\textsc{m}}\,\mathbf{v} \left(\mathbf{v}\cdot\mathbf{\hat{r}}_{\textsc{em}}\right)}{v_{\infty,\textsc{m}}^2 + u_{\mathrm{esc},\textsc{m}}^2 /2 - v_{\infty,\textsc{m}} \left(\mathbf{v}\cdot\mathbf{\hat{r}}_{\textsc{em}}\right)} \ .
\label{vinfm}
\end{equation}
Here, $v_{\infty,\textsc{m}}^2 = v^2 -  u_{\mathrm{esc},\textsc{m}}^2\,$, $u_{\mathrm{esc},\textsc{m}} = \sqrt{2GM_{\textsc{m}}/r_{\textsc{em}}(t)} \approx 0.1\ \mathrm{km/s}$, and $\mathbf{r}_{\textsc{em}}(t)$ is the position of the Earth in the Lunar frame. We note here that in reality, any particle which passes near the Moon is also necessarily affected by the gravitational pull of the Sun.  To account for this, we will make the approximation that the Sun's gravitational potential does not appreciably change in the vicinity of the Moon, so that we can take the incoming velocity of a particle in the Lunar frame to be given by the result of the Sun's gravitational deflection at the position of the Moon.\footnote{Additionally, one may worry that the Moon is accelerating with respect to the dark matter halo, and thus does not create a steady-state wake of dark matter as does the Sun which moves smoothly with respect the the dark matter halo.  Typical dark matter particles cross the gravitational well of the Moon in a matter of a few hours, during which the moon changes its position relative to the Earth by only a small fraction of Earth-Moon distance, and so the effect of the acceleration of the Moon can be safely neglected for our purposes.}

Putting all of this together then, the phase space density of dark matter in the Lab frame is given by
\begin{equation}
	\rho_{\chi} f(\mathbf{v},t)=\rho_{\infty}\tilde{f}\left(\mathbf{v}_{\odot}+v_{\infty,\textsc{s}}[\mathbf{v}_{\textsc{ms}}+ v_{\infty,\textsc{m}}[\mathbf{v}_{\textsc{em}}+\mathbf{v}]]\right) ,
\label{lville2}
\end{equation}
where $\mathbf{v}_{\textsc{ms}}$ is the velocity of the Moon in the Solar frame. Notice that the effects of annual and monthly motion of the Earth, as well as the the gravitational focusing due to the Sun and Moon have all been included here.

\section{Monthly Modulation}\label{monthly}

\subsection{WIMP Detection Experiments}\label{wimpdetection}
Dark matter direct-detection experiments seek to measure the recoil of nuclei in a detector due to scattering with dark matter particles.  The differential scattering rate for such events is given by 
\begin{equation}
	\frac{\diff R}{\diff E_{\mathrm{nr}}}=\frac{\rho_{\chi}}{2 m_{\chi} \mu^2}\sigma(q^2)\eta(v_{\mathrm{min}},t)\ ,
\label{diff rate}
\end{equation}
where $E_{\mathrm{nr}}$ is the nuclear recoil energy, $\rho_{\chi}$ is the local dark matter density, $m_{\chi}$ is the dark matter mass, $\mu$ is the reduced mass of the dark matter-nucleus system, $q$ is the momentum transfer, $\sigma(q^2)$ is the effective cross section of collision, and
\begin{equation}
	\eta(v_{\mathrm{min}},t)=\int_{v_{\mathrm{min}}}^\infty\ \! \frac{f(\mathbf{v},t)}{v} \, \Diff3v \ ,
\label{eta generic}
\end{equation}
where ${f}(\mathbf{v},t)$ is the dark matter velocity distribution in the Lab frame, and $v_\mathrm{min}$ is the minimum velocity required of an incoming dark matter particle to produce a recoil energy $E_{\mathrm{nr}}$.
We will assume in this section that the velocity distribution of WIMPs in the halo rest frame is given by the Standard Halo Model
\begin{equation}
 \tilde{f}(\mathbf{v}) =
  \begin{cases} 
      \frac{1}{N_{\mathrm{esc}}}\left(\frac{1}{\pi v_0^2}\right)^{3/2}\e^{-\mathbf{v}^2/v_0^2}\ , & |\mathbf{v}|<v_{\mathrm{esc}}\, , \\
      0\ , & \mathrm{else}\, ,
  \end{cases}
\label{fdist}
\end{equation}
where
\begin{equation}
N_{\mathrm{esc}}=\erf(z)-\frac{2}{\sqrt{\pi}}z\e^{-z^2}\, ,
\label{Nesc}
\end{equation}
and $z\equiv v_{\mathrm{esc}}/v_0$. We take $v_0=220$~km/s and set the escape velocity from the Galaxy to be $v_{\mathrm{esc}}=550$~km/s \cite{Lee:2013wza,Smith:2006ym}.  Small modifications to this velocity distribution should not strongly affect the conclusions of this section, though there are velocity distributions which can drastically affect modulation signatures \cite{Lee:2013xxa}.

\subsubsection{Motion of the Earth around the Earth-Moon Barycenter}\label{earth motion around bary}
The relationship between the velocity distribution of dark matter in the Galactic frame and the velocity distribution of dark matter in the Lab frame was given above by the Galilean transform  Eq. (\ref{boostedf}). To examine the monthly modulation in the detection signal, we set
\begin{equation}
	\mathbf{v}_{\mathrm{obs}}(t)=\mathbf{v}_{\odot}+\mathbf{v}_{\textsc{eb}}(t)+\mathbf{v}_{\textsc{bs}}(t) \, ,
	\label{vobs2}
\end{equation}
where $\mathbf{v}_{\textsc{bs}}$ is the velocity of the Earth-Moon barycenter around the Sun (see Appendix~\ref{coordinates} for the details). It is clear that the inclusion of the wobble of the Earth about the Earth-Moon barycenter in the definition of $\mathbf{v}_{\mathrm{obs}}$ will add some form of a monthly modulation to the differential rate, and we can estimate its size to be on the order $v_{\textsc{eb}}/v_{\textsc{bs}}\approx(1.3\times10^{-2} \,\mathrm{km/s})/(30 \, \mathrm{km/s})\approx 0.04 \%$ of the size of the annual modulation. Since the annual motion of the Earth around the Sun itself causes the rate to modulate by about $v_{\textsc{bs}}/(4v_{\odot}) \approx 3\%$ for $v_\textrm{min}\sim v_0$ \cite{Lewin:1995rx,Lee:2013xxa}, the monthly modulation should be approximately $10^{-5}$ times the size of the mean rate for similar detector thresholds. 
We now investigate this monthly effect in more detail.

If we neglect gravitational focusing, the function $\eta(v_\mathrm{min},t)$ defined in Eq.~(\ref{eta generic}) reads
\begin{equation}
\eta(v_{\mathrm{min}},t)=\int_{v_{\mathrm{min}}}^\infty\ \! \frac{\tilde{f}\left(\B{v}_\T{{obs}}(t) + \mathbf{v}\right)}{v} \, \Diff3v 
\equiv\eta(v_{\mathrm{min}},v_\mathrm{obs}(t))\, ,
\label{etanogf}
\end{equation}
where we have used the fact that for the Standard Halo Model in the absence of gravitational focusing, the time dependence of the mean inverse speed $\eta(v_\mathrm{min},t)$ enters only through the speed of the detector relative to the dark matter halo, $v_\mathrm{obs}(t)$. 
This integral can be evaluated analytically \cite{Lewin:1995rx,Freese:2003tt,Savage:2006qr} to obtain
\begin{equation}
 \eta(v_{\mathrm{min}},v_\mathrm{obs}(t)) =
  \begin{cases} 
      \frac{1}{2N_{\mathrm{esc}}yv_0}\left[\erf(x+y)-\erf(x-y)-\frac{4}{\sqrt{\pi}}y\e^{-z^2}\right]\ , & x<z-y\, , \\[0.5em]
      \frac{1}{2N_{\mathrm{esc}}yv_0}\left[\erf(z)-\erf(x-y)-\frac{2(y+z-x)}{\sqrt{\pi}}\e^{-z^2}\right]\ , & z-y<x<z+y\, , \\[0.5em]
			0\ , & x>z+y\, ,
  \end{cases}
\label{etaanalytic}
\end{equation} 
where we have defined  $x\equiv v_{\mathrm{min}}/v_0$ and $y\equiv v_{\mathrm{obs}}(t)/v_0\,$; recall that $z\equiv v_{\mathrm{esc}}/v_0$. 

\begin{figure}[h]%
\centering
\includegraphics[width=0.8\textwidth]{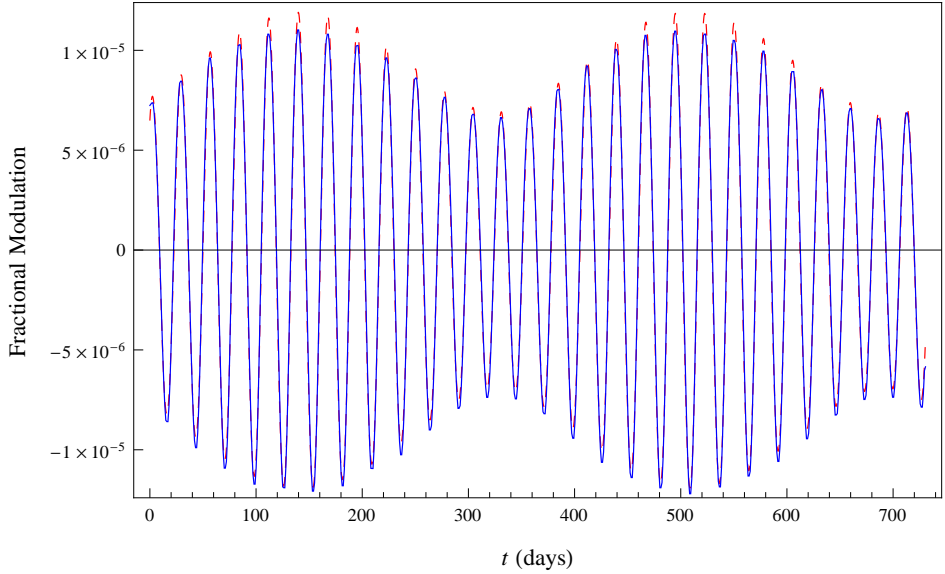}%
\caption{Plotted in blue is the fractional modulation over two years (starting from J2000.0), for $v_\T{min}=100 \,\T{km/s}$. The fractional modulation is defined as $\Delta\eta/\langle\eta\rangle$ (with $\Delta\eta$ itself defined in Eq.~(\ref{etares})). The dashed red line is the plot of the fractional modulation using the approximation to $\Delta\eta$ given in Eq.~(\ref{etaresfull}).}%
\label{baryvmin100fig}%
\end{figure}

In order to isolate the effects of the Earth's motion about the Earth-Moon barycenter, we subtract from Eq.~(\ref{etanogf}) the effects of the annual motion of the Earth around the Sun, which can be obtained by computing $\eta$ without the barycentric wobble of the Earth:
\begin{equation}
\Delta\eta(v_{\mathrm{min}},t)=\eta(v_{\mathrm{min}},|\mathbf{v}_{\odot}+\mathbf{v}_{\textsc{es}}(t)|)-\eta(v_{\mathrm{min}},|\mathbf{v}_{\odot}+\mathbf{v}_{\textsc{bs}}(t)|)\ .
\label{etares}
\end{equation}
Fig.~\ref{baryvmin100fig} shows a plot of this residual function at $v_{\mathrm{min}}=100\ \mathrm{km/s}$. Note that for the modulation plots throughout this work, we will plot the dimensionless ``fractional modulation'' on the $y$-axis (defined as $\Delta\eta/\langle\eta\rangle$, where angle brackets refer to the time average) to indicate the size of the modulation as compared to the mean rate, and time measured in days from J2000.0 on the $x$-axis.

First, notice from the figure that the estimate we made above of the relative size of the monthly modulation to the annual is in fact a good one. It is clear however, that the effect of the Earth's wobble is more complicated than a simple monthly modulation: the amplitude of the monthly modulation itself modulates annually.

In order to explain this behavior, we will treat the velocity of the Earth relative to the Earth-Moon barycenter $\mathbf{v}_{\textsc{eb}}(t)$ as a small perturbation to $\mathbf{v}_{\mathrm{obs}}(t)$ in $\eta$.  Using the spherical symmetry of the distribution function, the Taylor expansion of $\eta$ reads
\begin{equation}
\eta\left(v_{\mathrm{min}},\left|\tilde{\mathbf{v}}_\mathrm{obs}+\delta\mathbf{v}_\mathrm{obs}\right|\right)=\eta\left(v_{\mathrm{min}},\left|\tilde{\mathbf{v}}_\mathrm{obs}\right|\right)+\left.\frac{\partial\eta\left(v_{\mathrm{min}},v_\mathrm{obs}\right)}{\partial v_{\mathrm{obs}}}\right|_{v_\mathrm{obs}=\tilde{v}_{\mathrm{obs}}}\left(\hat{\tilde{\mathbf{v}}}_{\mathrm{obs}} \cdot \delta\mathbf{v}_{\mathrm{obs}}\right)+\mathcal{O}(\delta v_{\mathrm{obs}}^{\,2}) \ .
\label{etataylor}
\end{equation}
We can hence approximate the residual function Eq.~(\ref{etares}) up to terms of order $v_\textsc{eb}^2$ as
\begin{equation}
\Delta\eta(v_{\mathrm{min}},t)\simeq\left.\frac{\partial\eta\left(v_{\mathrm{min}},v_\mathrm{obs}\right)}{\partial v_\mathrm{obs}}\right|_{v_\mathrm{obs}=|\mathbf{v}_\odot+\mathbf{v}_\textsc{bs}(t)|}\left(\frac{\mathbf{v}_\odot+\mathbf{v}_\textsc{bs}(t)}{|\mathbf{v}_\odot+\mathbf{v}_\textsc{bs}(t)|} \cdot \mathbf{v}_\textsc{eb}(t)\right)\ ,
\label{etaresfull}
\end{equation}
where we have made the identifications $\tilde{\mathbf{v}}_{\mathrm{obs}}(t)=\mathbf{v}_{\odot}+\mathbf{v}_{\textsc{bs}}(t)$ and $\delta\mathbf{v}_{\mathrm{obs}}(t)=\mathbf{v}_{\textsc{eb}}(t)$. Fig~\ref{baryvmin100fig} shows that this approximation provides an excellent fit to $\Delta\eta$.

Since Eq.~(\ref{etaresfull}) is such a good approximation to $\Delta\eta$, it is worth examining it in some detail. We shall see presently that it analytically describes all of the features of $\Delta\eta$, both as a function of $v_{\mathrm{min}}$ and $t$.

\begin{figure}[tbh]%
\centering
\includegraphics[width=0.8\textwidth]{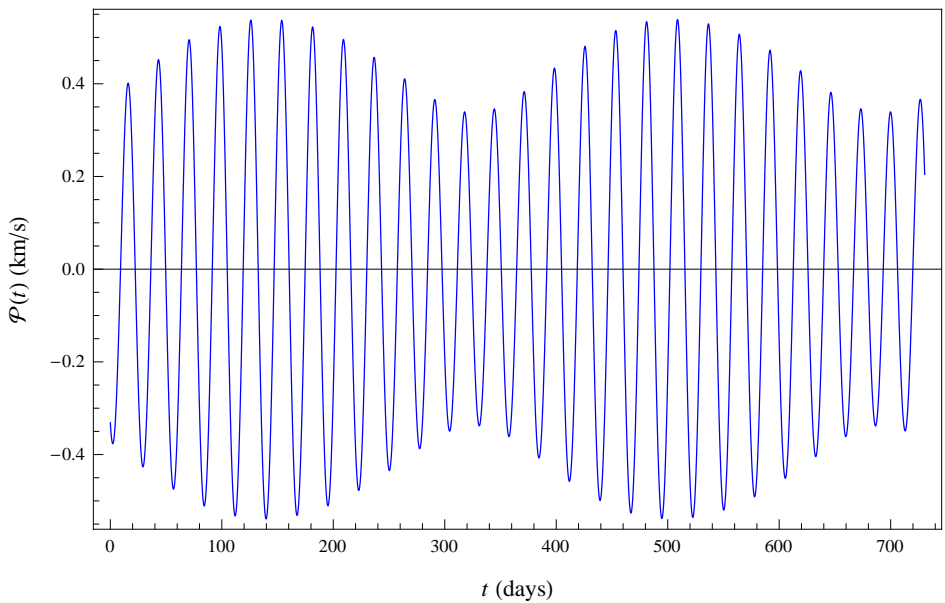}%
\caption{Plot showing the time dependence of $\mathcal{P}(t)$, defined in Eq.~(\ref{omegasyn2}).  Notice that due to the presence of terms with both the sidereal and synodic monthly periods, $\mathcal{P}(t)$ exhibits an annual ``beat'' which then also appears in $\Delta\eta$; see Fig.~\ref{baryvmin100fig} for comparison.}%
\label{dotproductfig}%
\end{figure}

Consider then the time dependence of the dot product appearing in Eq.~(\ref{etaresfull}) (this is plotted in Fig.~\ref{dotproductfig}). The unit vectors $\hat{\mathbf{v}}_{\textsc{bs}}$ and $\hat{\mathbf{v}}_{\textsc{eb}}$ to zeroth order in eccentricity take the form
\begin{align}
	\hat{\mathbf{v}}_{\textsc{bs}}(t) &= \boldsymbol{\hat{\epsilon}}_{1}\cos{(\omega_{\mathrm{yr}}t-\phi_1)} + \boldsymbol{\hat{\epsilon}}_{2}\sin{(\omega_{\mathrm{yr}}t-\phi_1)}\, ,\\
	\hat{\mathbf{v}}_{\textsc{eb}}(t) &= \boldsymbol{\hat{\epsilon}}_{1,\textsc{m}}\cos{(\omega_{\mathrm{sid}}t-\phi_2)} + \boldsymbol{\hat{\epsilon}}_{2,\textsc{m}}\sin{(\omega_{\mathrm{sid}}t-\phi_2)}\, ,
\label{velocityexpansions}
\end{align}
where $\omega_{\mathrm{yr}}=2\pi/T_{\mathrm{yr}}$ and $\omega_{\mathrm{sid}}=2\pi/T_{\mathrm{sid}}$ are the angular frequencies associated with the periods of the orbits of the Earth around the Sun, and the Moon around the Earth (as measured against the celestial sphere), respectively. We will leave the phases unspecified in these intermediate steps for clarity, but it is straightforward to retain them throughout the calculation. The dot product of these unit vectors is given by
\begin{equation}
	\hat{\mathbf{v}}_{\textsc{bs}}(t)\cdot\hat{\mathbf{v}}_{\textsc{eb}}(t)= \frac{1+\cos i_\textsc{m}}{2}\cos{((\omega_{\mathrm{sid}}-\omega_{\mathrm{yr}})t-\phi_3)}+\frac{1-\cos i_\textsc{m}}{2}\cos{((\omega_{\mathrm{sid}}+\omega_{\mathrm{yr}})t-\phi_4)}\, ,
\label{omegasyn}
\end{equation}
where $i_\textsc{m}\approx 5.2^\circ$ is the inclination of the orbital plane of the Moon with respect to the ecliptic.  We approximate $\cos i_\textsc{m}\approx1$ for simplicity here.  Recognizing the difference between the sidereal and annual frequencies as precisely the synodic frequency, $\omega_{\mathrm{syn}}\equiv \omega_{\mathrm{sid}}-\omega_{\mathrm{yr}}$, we obtain
\begin{equation}
	\hat{\mathbf{v}}_{\textsc{bs}}(t)\cdot\hat{\mathbf{v}}_{\textsc{eb}}(t)\simeq \cos{(\omega_\mathrm{syn}t-\phi_3)} \, .
\end{equation}
We can now use this result to compute the dot product appearing in Eq.~(\ref{etaresfull}):
\begin{align}
	\mathcal{P}(t) & \equiv \left(\frac{\mathbf{v}_\odot+\mathbf{v}_\textsc{bs}(t)}{|\mathbf{v}_\odot+\mathbf{v}_\textsc{bs}(t)|} \cdot \mathbf{v}_\textsc{eb}(t)\right) \nonumber \\
	&\simeq\frac{v_\textsc{eb}}{|\mathbf{v}_\odot+\mathbf{v}_\textsc{bs}(t)|}\left[v_\odot b_\textsc{m} \cos{(\omega_\textrm{sid}(t-t_1))} +v_\textsc{bs}\cos{(\omega_\mathrm{syn}(t-t_2))}\right] \nonumber \\
	&=\frac{v_\textsc{eb}}{|\mathbf{v}_\odot+\mathbf{v}_\textsc{bs}(t)|}\Bigg[\left(v_\odot b_\textsc{m}-v_\textsc{bs}\right)\cos{(\omega_\textrm{sid}(t-t_1))} \nonumber \\
	&\qquad\qquad+2v_\textsc{bs}\cos{\left(\frac{\omega_\mathrm{syn}+\omega_\mathrm{sid}}{2}t-\phi_a\right)}\cos{\left(\frac{\omega_\mathrm{yr}}{2}t-\phi_b\right)}\Bigg] \, ,
\label{omegasyn2}
\end{align}
where $t_1$ is the time when $\mathbf{v}_\textsc{eb}$ is most nearly parallel to $\mathbf{v}_\odot$, and $t_2$ is the time when $\mathbf{v}_\textsc{eb}$ is most nearly parallel to $\mathbf{v}_\textsc{bs}$. We have further defined the geometric factor $b_\textsc{m}\equiv\sqrt{\left(\hat{\mathbf{v}}_\odot \cdot \boldsymbol{\hat{\epsilon}}_{1,\textsc{m}}\right)^2+\left(\hat{\mathbf{v}}_\odot \cdot \boldsymbol{\hat{\epsilon}}_{2,\textsc{m}}\right)^2}\simeq0.45$, which accounts for the alignment of the Moon's orbital plane with the motion of the Sun. The synodic period $T_{\T{syn}}\approx 29.53\ \T{days}$ is the average period of the Moon's revolution with respect to the line joining the Sun and Earth, and thus determines the period of the moon's phases. From this equation therefore, we see that $\mathcal{P}(t)$ depends both on the synodic and the sidereal period of the Earth's wobble about the Earth-Moon barycenter, and it is the presence of both these frequencies in the residual function produces that an annual ``beat'' in the differential scattering rate. 

\begin{figure}[htb]%
\centering
\includegraphics[width=0.9\textwidth]{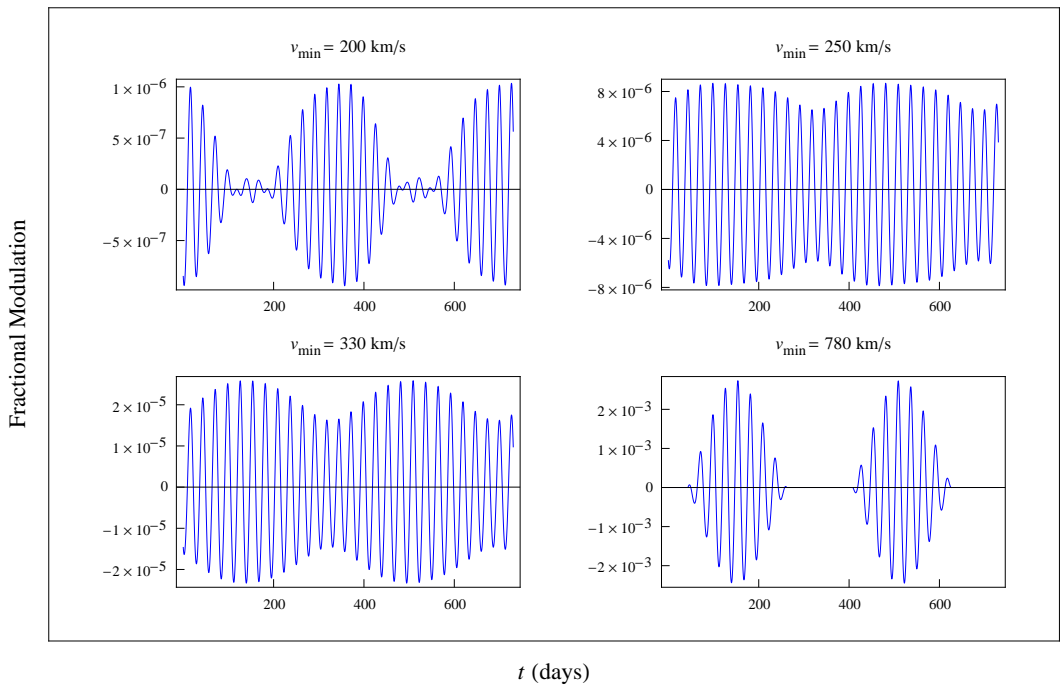}%
\caption{We plot here the fractional modulation $\Delta\eta/\langle\eta\rangle$ for various values of $v_\T{min}$. The increasing amplitude on the $y$-axis for larger values of $v_\mathrm{min}$ is primarily due to the decrease in $\langle\eta\rangle$ (see Fig.~\ref{meanratefig}).  The changes in shape come from the $v_\mathrm{min}$ dependence of the function $\mathcal{A}$ (see Fig.~\ref{aslice}).}%
\label{barycomposite}%
\end{figure}

\begin{figure}[htb]%
\centering
\includegraphics[width=0.5\textwidth]{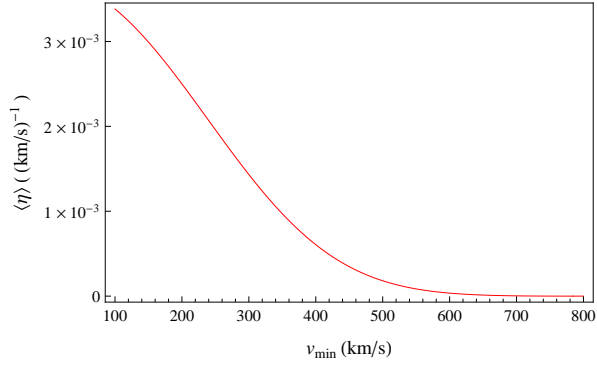}%
\caption{This plot shows the time-averaged mean inverse speed $\langle\eta\rangle$ as a function of $v_\T{min}$; notice that it is a monotonically decreasing function of $v_\mathrm{min}$ and that it drops to zero for $v_\mathrm{min}\GtrSim800$~km/s.}%
\label{meanratefig}%
\end{figure}

\begin{figure}[htb]
        \centering
        \begin{subfigure}{0.45\textwidth}
                \includegraphics[width=\textwidth]{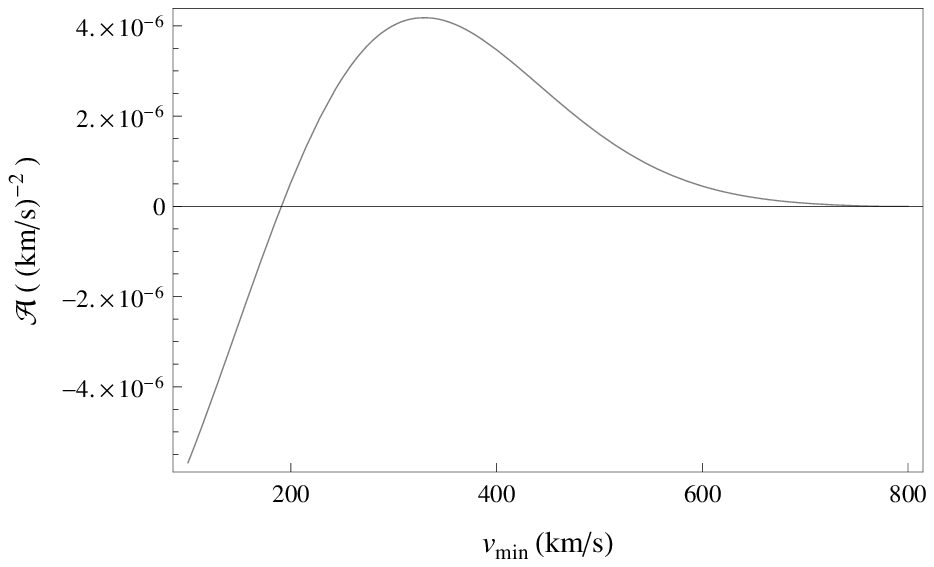}
                \caption{A slice of $\mathcal{A}$ along $t=0$, showing the $v_\mathrm{min}$ dependence.}
                \label{aslicet}
        \end{subfigure}%
        ~ 
        \begin{subfigure}{0.45\textwidth}
                \includegraphics[width=\textwidth]{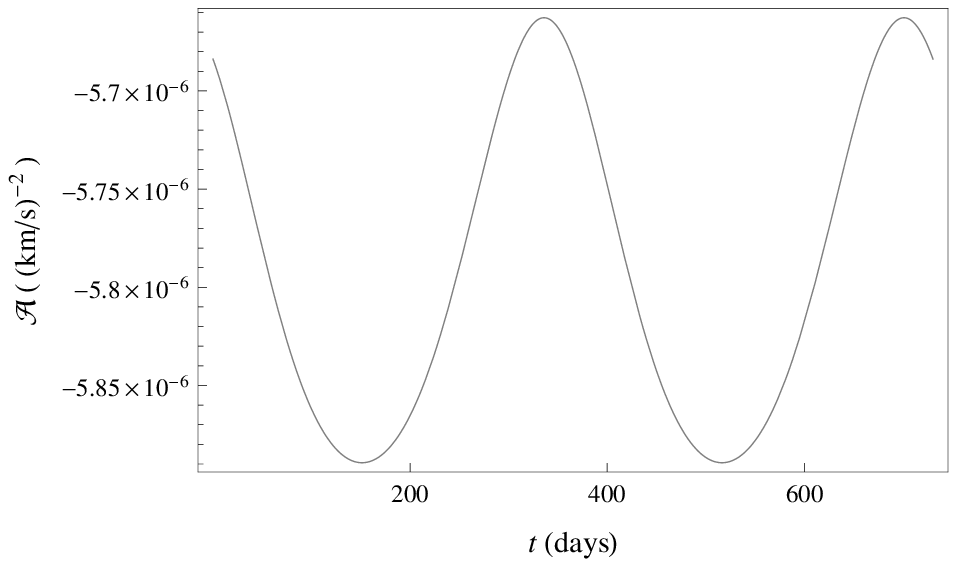}
                \caption{A slice of $\mathcal{A}$ along $v_\T{min}=100$~km/s, showing the time dependence.}
                \label{aslicevmin}
        \end{subfigure}       
        \caption{Plots showing the function $\mathcal{A}(v_{\T{min}},t)$ defined in Eq.~(\ref{A(vmin)}).}
				\label{aslice}
\end{figure}

We turn now to the functional dependence of the fractional modulation $\Delta\eta$ on $v_{\T{min}}$, which is directly related to detector energy threshold. In Fig.~\ref{barycomposite} we plot $\Delta\eta$ for four different values of $v_\textrm{min}$.  Note that both $\Delta\eta$ and $\langle\eta\rangle$ are functions of the threshold speed $v_\mathrm{min}$. $\langle\eta\rangle$ is a monotonically decreasing function of $v_\mathrm{min}$ (see Fig.~\ref{meanratefig}), while the behavior of $\Delta\eta$ is more complicated and is described by the function
\begin{equation}
\mathcal{A}(v_{\T{min}},t)\equiv \left.\frac{\partial\eta(v_{\mathrm{min}},v_\mathrm{obs})}{\partial v_\mathrm{obs}}\right|_{v_\mathrm{obs}=|\mathbf{v}_\odot+\mathbf{v}_\textsc{bs}(t)|}\,  ,
\label{A(vmin)}
\end{equation}
which is the other piece of the approximation Eq. (\ref{etaresfull}) above. It can be computed analytically:
\begin{equation}
	\mathcal{A}(v_{\T{min}},t)=
		\begin{cases} 
      \frac{1}{2N_{\mathrm{esc}}y^2v_0^2}\left[-\erf(x+y)+\erf(x-y)+\frac{2}{\sqrt{\pi}}y\left(\e^{-(x+y)^2}+\e^{-(x-y)^2}\right)\right]\ , & x<z-y\, , \\[0.5em]
      \frac{1}{2N_{\mathrm{esc}}y^2v_0^2}\left[-\erf(z)+\erf(x-y)-\frac{2}{\sqrt{\pi}}\left((z-x)\e^{-z^2}+y\e^{-(x-y)^2}\right)\right]\ , & z-y<x<z+y\, , \\[0.5em]
			0\ , & x>z+y\, .
  \end{cases}
\end{equation}
See Fig.~\ref{aslice} for plots of this function. 

First, consider the amplitude of the fractional modulation shown in Fig.~\ref{baryvmin100fig} and Fig.~\ref{barycomposite}. We see (apart from a decrease in amplitude between $v_\mathrm{min}=100$~km/s and $v_\mathrm{min}=200$~km/s) that the size of fractional modulation increases with $v_\T{min}$, which might seem to imply that modulation is most easily detectable for experiments with high detector threshold. But since the mean rate decreases quite rapidly with $v_\mathrm{min}$ (Fig.~\ref{meanratefig}), the small mean rate in this range makes any detection of dark matter more difficult than for experiments with lower thresholds, despite the large fractional modulation at high $v_\T{min}$.

Another aspect of $\mathcal{A}$ that is important in describing features of $\Delta\eta$ is that for a fixed $v_{\T{min}}$, $\mathcal{A}$ modulates annually with $t$; see Fig.~\ref{aslicevmin}. Note that this effect is distinct from, and competes with, the annual beat in $\mathcal{P}$, shown earlier in Fig.~\ref{dotproductfig}. For instance, the annual envelope in $\Delta\eta$ for $v_{\T{min}}$ in the range $220-300\ \T{km/s}$ is muted compared to the envelope for $\mathcal{P}$, because the annual feature in $\mathcal{A}$ is out of phase with the annual envelope in $\mathcal{P}$ in this range. Then for $v_\T{min}$ in the range $300-330\ \T{km/s}$, the envelope looks much closer to that of $\mathcal{P}$, since the annual feature of $\mathcal{A}$ flattens out for these values. Finally, for $v_{\T{min}}\ge 330\ \T{km/s}$, the characteristics of the envelope in $\Delta\eta$ are pronounced, owing to the fact that the peaks and troughs of $\mathcal{A}$ are now in phase with those of $\mathcal{P}$.

Additionally, because $\mathcal{A}$ is negative for $v_{\T{min}}\LessSim 200\ \T{km/s}$, the fractional modulation has a phase opposite that of $\mathcal{P}$ in this range; compare Figs.~\ref{baryvmin100fig} and \ref{dotproductfig}. In fact, this flip causes the annual components of $\mathcal{A}$ and $\mathcal{P}$ to line up, accentuating the annual envelope in $\Delta\eta$. The crossing of $\mathcal{A}$ from negative to positive at $v_{\mathrm{min}}\approx 200\ \T{km/s}$ causes a shift in the peak of the annual envelope from early December to early June, which in turn results in the peak of the monthly modulation shifting by about two weeks; this can be used to determine the dark matter mass \cite{Lewis:2003bv}.

\subsubsection{Gravitational Focusing by the Moon}

The gravitational well of the Moon distorts the local distribution of dark matter and introduces an additional source of modulation in the rate count of direct-detection experiments. In this section, we examine this focusing effect. Recalling Eq.\ (\ref{lville2}) which related the Lab frame and Galactic frame velocity distributions of dark matter, and putting it together with the expression for the mean inverse speed Eq.\ (\ref{eta generic}), we obtain the following:
\begin{equation}
	\rho_\chi\int_{v_{\mathrm{min}}}^\infty\ \! \frac{f\left(\mathbf{v},t\right)}{v} \,\Diff3v =
	\rho_\infty\int_{v_{\mathrm{min}}}^\infty\ \! \frac{\tilde{f}\left(\mathbf{v}_{\odot}+v_{\infty,\textsc{s}}[\mathbf{v}_{\textsc{ms}}+ v_{\infty,\textsc{m}}[\mathbf{v}_{\textsc{em}}+\mathbf{v}]]\right)}{v} \,\Diff3v \ .
\label{eta}
\end{equation}
Here, $\mathbf{v}_{\textsc{ms}}$ is the velocity of the Moon in the Solar frame. Notice that the effects of annual and monthly motion of the Earth, as well as the the gravitational focusing due to the Sun and Moon have been included here.

To estimate the effects of gravitational focusing, we first note that focusing effects peak when the detector is positioned most nearly behind the focusing body with respect to the stream of incoming dark matter particles, and this occurs a quarter period before (or after) the peak of the modulation due to the motion of the detector, which itself occurs when the detector moves most nearly toward (or away from) the stream of incoming particles.  Said another way, the two effects peak at different times, so it is their relative size which determines the position of the peak in the actual rate count observed at a detector. In the case of the annual modulation, since the size of the effect of the Sun's gravitational focusing is similar in magnitude to the effect of the Earth's motion around the Sun (for low detector thresholds), gravitational focusing results in a phase shift of the annual modulation \cite{Lee:2013wza}. Hence, in order to determine whether a similar effect exists for the monthly modulation, we need to estimate the magnitude of the effect of the gravitational focusing effect due to the Moon, and compare it to the size of the effect of the Earth's barycentric wobble.

The modulation due to the focusing of the Moon scales roughly as $(u_{\T{esc},m}/v)^2$ \cite{Lee:2013wza}, and so for $v=300\ \T{km/s}$, we expect a monthly modulation on the order of $10^{-7}\%$ of the mean rate due to the gravitational focusing of the Moon alone. We have already estimated that the relative size of the barycentric effect to the annual is $v_\textsc{eb}/v_\textsc{bs}\approx 0.04\%$, and as stated above, the annual motion of the Earth around the Sun causes the mean rate to modulate by about $v_{\textsc{bs}}/(4v_{\odot}) \approx 3\%$. Therefore, we see that relative size of the effect of the Moon's gravitational focusing to the barycentric motion is approximately
\begin{equation}
	\left(\frac{u_{\T{esc},m}}{v}\right)^2/\left(\frac{v_\textsc{eb}}{v_\textsc{bs}}\frac{v_\textsc{bs}}{4v_\odot}\right)=0.008\%\ ,
\label{estimate}
\end{equation}
and so we expect a negligible modification to the monthly modulation as a result of the Moon's gravitational focusing.

To confirm this estimate, we carried out the full computation as follows. Since the integral appearing in Eq.~(\ref{eta}) cannot be evaluated analytically, we computed it numerically, and studied the function
\begin{equation}
\Delta\eta_{\,\textsc{m}}(v_{\T{min}},t) \equiv \int_{v_{\mathrm{min}}}^\infty\ \! \left[ \frac{\tilde{f}\left(\mathbf{v}_{\odot}+v_{\infty,\textsc{s}}[\mathbf{v}_{\textsc{ms}}+ v_{\infty,\textsc{m}}[\mathbf{v}_{\textsc{em}}+\mathbf{v}]]\right)}{v} -  \frac{\tilde{f}\left(\mathbf{v}_{\odot}+v_{\infty,\textsc{s}}[\mathbf{v}_{\textsc{bs}}+\mathbf{v}]\right)}{v}\right] \,\Diff3v\ , 
\label{delta eta moon}
\end{equation}
where the second term includes only the effects of annual modulation compounded by the Sun's gravitational focusing. The function $\Delta\eta_{\,\textsc{m}}$ then, is a residual of all monthly signatures: it includes the effects of the barycentric motion of the Earth and the gravitational focusing of the Moon, and any compound effects.

\begin{figure}[h]%
\centering
\includegraphics[width=0.8\textwidth]{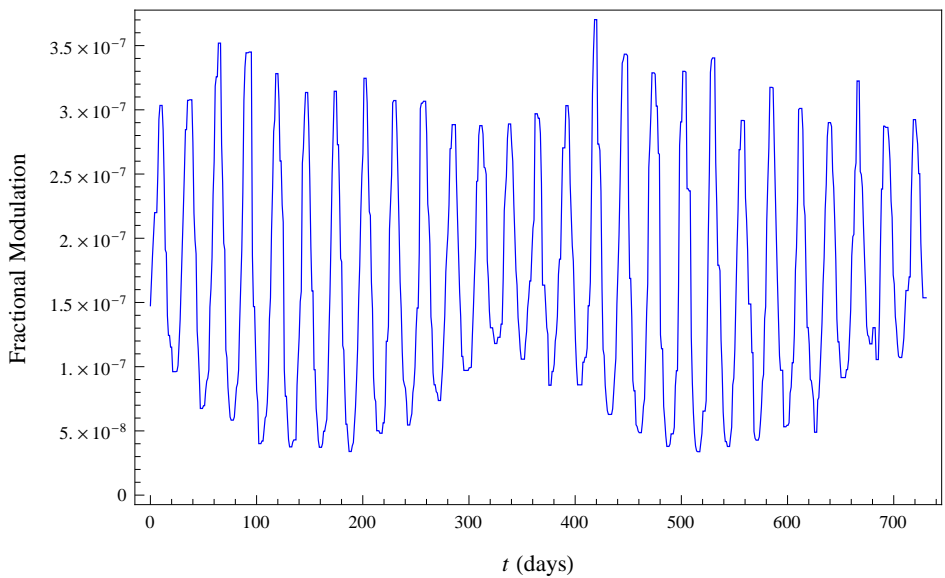}%
\caption{Isolating the effect of the Moon's gravitational focusing, this plot shows $\left(\Delta\eta_{\,\textsc{m}}-\Delta\eta\right)/\langle\eta\rangle$ at $v_\T{min}=100\ \T{km/s}$ over two years.  The amplitude is even smaller for larger values of $v_\mathrm{min}$.  Some numerical noise has been introduced by the integration methods used to construct this plot.}%
\label{moonfocusingfig}%
\end{figure}

We fit $\Delta\eta_{\,\textsc{m}}$ for various $v_\T{min}$ and $t$ with functions of the form
\begin{equation}
\Delta\eta_{\,\textsc{m},\textrm{fit}}=C+A\cos{(\omega_{\T{sid}}t-\phi_\textsc{a})}+B\cos{(\omega_{\T{syn}}t-\phi_\textsc{b})} \ \,
\label{eq:monthlyfit}
\end{equation}
and compared the amplitudes, $A$ and $B$, and the phases, $\phi_\textsc{a}$ and $\phi_\textsc{b}$, with those from analogous fits made to $\Delta\eta$ from Eq.~(\ref{etares}). We found that the parameters of the fit were essentially unchanged, confirming that the gravitational focusing of the Moon has a negligible impact on the monthly modulation from the barycentric wobble. One can further see this by computing the function $\Delta\eta_{\,\textsc{m}}-\Delta\eta$, which isolates the effect of the Moon's gravitational focusing on the rate; see Fig.~\ref{moonfocusingfig} for a plot. Notice that the graph shares several features with that of $\Delta\eta$, most noticeably the annual envelope, which arises once again from the interaction between $\omega_\T{sid}$ and $\omega_\T{syn}$. More significant is the fact that the fractional modulation of $\Delta\eta_{\,\textsc{m}}-\Delta\eta$, even though larger for $v_\mathrm{min}=100$~km/s than the estimate made above, is significantly smaller than the fractional modulation for $\Delta\eta$ itself.  Further, the effect of gravitational focusing decreases with increasing threshold speed, and so it is even less important for higher $v_\mathrm{min}$.  For all practical purposes then, the gravitational focusing due to the Moon can be ignored when considering monthly modulations in the rate.

In addition to $\Delta\eta_{\,\textsc{m}}$, several other residual functions were computed and examined to check for interplay between the various effects discussed above. For instance, we studied the interaction between the gravitational focusing due to the Sun and the annual envelope of the barycentric wobble signature. The findings from these computations can be summarized simply as follows: (a) the annual modulation compounded by the effects of the Sun's gravitational focusing is the most dominant feature in the rate, and it is essentially unaffected by the addition of the barycentric wobble to the velocity, and (b) the monthly modulation due to the Earth's motion about the Earth-Moon barycenter, as seen most directly by computing the residual function $\Delta\eta$, is not significantly affected by focusing from either the Sun or the Moon.

\subsection{WISP Detection Experiments}

Direct searches for WISP dark matter seek to observe the conversion of a WISP particle into a photon inside a detector.  The energy of the resulting photon is determined by energy conservation to be
\begin{equation}
	hf=m c^2\left(1+\frac{1}{2}\frac{v^2}{c^2}\right)\, ,
\end{equation}
where $m$ is the mass of the WISP particle, and $v$ is the speed of the particle relative to the detector.  This conversion can be made more efficient in a resonant cavity when the frequency of the photons produced matches the resonant frequency of the cavity \cite{Sikivie:1983ip}.  This is the strategy pursued for example by ADMX \cite{Asztalos:2001tf,Asztalos:2009yp}.  The frequency of the signal from WISP to photon conversion is measured much more accurately than the power, and so we will focus our attention on the time dependence of the observed frequency.

For WISPs moving at a fixed velocity relative to the Sun, changes in the detector velocity due to the motion of the Earth result in modulation of the observed photon frequency. The shift in frequency $\Delta f$ due to a small change in the velocity of the detector relative to the incoming WISPs $\Delta v$ is given by
\begin{equation}
	\Delta f = \frac{f_0 v \Delta v}{c^2} \, ,
\end{equation}
where $f_0=\frac{mc^2}{h}$. For WISPs with mass around 2 $\mu$eV, we have $f_0\approx 500$ MHz, which is within the search window of ADMX \cite{Asztalos:2001tf,Asztalos:2009yp}.

In section \ref{wimpdetection} above, we made a rather simple assumption about the dark matter velocity distribution.  However, since WISP dark matter searches are more sensitive to the velocity of incoming dark matter particles, it is useful to relax this assumption here.  The velocity distribution of our dark matter halo may contain components with very low velocity dispersion, as is expected for example in the caustic ring model \cite{Sikivie:1997ng,Sikivie:1999jv} or due to late infall of dark matter \cite{Sikivie:1992bk,Natarajan:2005fh}.  These cold flows typically have a very small spread of velocities with a mean ranging from about 100 km/s to about 1000 km/s. In the event that the dark matter velocity distribution contains such low velocity dispersion components, a higher signal-to-noise can be achieved in experiments like ADMX by scanning over narrower frequency bins \cite{Duffy:2005ab,Duffy:2006aa,Hoskins:2011iv}.  High resolution searches of this type are currently underway at ADMX, and the maximum resolution presently utilized in ADMX has a bin size of 19 mHz.

The daily rotation of the Earth and its annual orbit around the Sun result in frequency shifts which must be taken into account in experiments like the high resolution search of ADMX \cite{Ling:2004aj}.  The monthly motion of the Earth about the Earth-Moon barycenter also results in a modulation of the expected photon frequency which cannot be neglected in general.  For example, a cold flow of 2 $\mu$eV WISPs with velocity 600 km/s directed in the plane of the Moon's orbit will produce photons with a frequency which differs by 87 mHz compared to those produced two weeks later (neglecting here the annual and daily motion of the Earth).  Since this frequency shift is larger than the minimum bin size that is utilized in ADMX, it must in general be taken into account when analyzing the data, especially when combining data collected days or weeks apart.

Gravitational focusing by the sun results in an annual modulation of the signal frequency for WISP searches \cite{Ling:2004aj}, and gravitational focusing by the Moon also contributes to a monthly modulation of the signal frequency.  If the Moon were at a constant distance from the Earth, all incoming dark matter particles would have their kinetic energy increased by the same amount due to the gravitational attraction of the Moon.  This would create a constant upward shift in the frequencies of all photons detected in a WISP search.  Due to the eccentricity of the Moon's orbit, however, there is a monthly periodicity to this effect, leading to a difference of about 9 m/s in the speed of incoming WISPs when the Moon is at apogee compared to when it is at perigee.  For 2 $\mu$eV WISPs with a velocity of 600 km/s, this leads to a shift of about 30 mHz in the detected frequency, which is also larger than the minimum bin size used in the ADMX high resolution search.  As a result, both the time-dependent distance to the Moon and the motion of the Earth about the Earth-Moon barycenter must be taken into account in order to properly treat the monthly modulation of WISP signal frequencies.

If the dark matter halo of our galaxy is well described by a smooth velocity distribution, like the Standard Halo Model defined in Eq.~(\ref{fdist}), it will be more difficult to detect the effects of monthly modulation.  In this case, the WISP signal would be non-vanishing over a frequency range which is quite broad compared to the frequency shifts due to modulation.  For 2 $\mu$eV WISP dark matter described by the Standard Halo Model, the signal would smoothly vary over a range of about 700 Hz, while the frequency shifts due to the motion of the Earth about the Earth-Moon barycenter and the gravitational focusing of the moon would lead to shifts of the entire signal by tens of mHz with a monthly period.  The fractional change in any given frequency bin due to monthly modulation would be on the order of $10^{-4}$.  While this shift is detectable in principle, it would be very difficult to do so in practice.  For this reason, if our dark matter halo has a velocity distribution with some non-trivial features, it would allow a much easier detection of modulation effects.

Just as in the case of WIMP searches, the modulation of candidate WISP signal frequency could be used to distinguish such events from background, which would not be expected to undergo the same modulation.  Furthermore, if we are fortunate enough to detect WISP dark matter in an experiment like ADMX, it would be very interesting to follow up such a detection by making a detailed study of the local velocity distribution of dark matter.  A proper mapping of detected photon frequencies to dark matter velocities requires a careful accounting of the Earth's motion, and combining the daily, monthly, and annual modulation of frequencies would allow a determination of the components of dark matter velocity lying along the plane of the Earth's rotation, Moon's orbit, and Earth's orbit, respectively.

\section{Discussion and Conclusion}\label{discussion}

Modulation in dark matter direct-detection experiments is generically expected due to the motion of the Earth and the gravitational effects of the Sun, Earth, and Moon.  Study of the modulation provides a useful tool for distinguishing signal events from background for both WIMP and WISP searches.  Annual modulation is the most prominent and readily detectable type of modulation.  However, there are several potential sources of background which also experience annual modulation, which motivates a deeper understanding of the expected time dependence of dark matter signals.

Let us now briefly examine diurnal modulation for WIMP searches, which we have so far ignored.  The rotation of the Earth imparts a velocity to the detector with magnitude $v_\mathrm{rot}\approx0.46\cos{\varphi_0}$~km/s relative to the center of the Earth, where $\varphi_0$ is the geographical latitude of the detector.  This motion results in a daily modulation of the dark matter scattering rate which is approximately $(2.2\cos{\varphi_0})\,\%$ of the annual modulation, or about $(0.066\cos{\varphi_0})\,\%$ of the mean rate \cite{Lee:2013xxa}.\footnote{Following an analysis similar to that in Sec.~\ref{earth motion around bary} shows that both the sidereal and synodic daily periods of the Earth's rotation will contribute to the diurnal modulation, leading to an annual cycle in the amplitude of the diurnal modulation.} On the other hand, the effect of gravitational focusing due to the Earth is more complicated than for the Sun or Moon: for a detector near the surface of the Earth, many particles will have passed through the bulk of the Earth before arriving at the detector, and a formula like Eq.~(\ref{vinfs}) is no longer sufficient to calculate the effect of the Earth's gravity on each particle's velocity; energy conservation still dictates that $v_{\infty,\textsc{e}}^2=v^2-u_{\mathrm{esc},\textsc{e}}^2$, but the angular dependence of the focusing will be complicated.  Despite these complexities, we can naively estimate the size of the focusing effect to be $(u_{\mathrm{esc},\textsc{e}}/v)^2\approx0.14\%$ of the mean rate for particles travelling at 300~km/s, which is more than twice as large as the effect of the rotational speed of the Earth.  Additionally, the eclipsing of the incoming flux of dark matter particles by the bulk of the Earth (separate from the focusing effect just mentioned) contributes to diurnal modulation, and the size of this effect depends upon the dark matter properties as well as the geographical location of the detector \cite{Collar:1992qc,Hasenbalg:1997hs,Kouvaris:2014lpa,Foot:2014osa}. Since the diurnal modulation of dark matter scattering rate results from a combination of these three effects, making definite predictions for the expected modulation is challenging.  In addition, there are daily cycles which affect potential sources of background. Therefore, despite the larger expected amplitude of daily modulation, a detection of monthly modulation in a WIMP dark matter experiment would provide more conclusive evidence in favor of dark matter than would daily modulation.

The DAMA experiment has observed an annual modulation whose magnitude is $(0.0112 \pm 0.0012)$~cpd/kg/keV \cite{Bernabei:2013xsa}.  If this modulation is indeed due to dark matter, one should expect a monthly modulation with an amplitude of roughly $4\times10^{-6}$~cpd/kg/keV, which is unfortunately far below the current sensitivity of the experiment.  Thus, it seems as though very significant improvements in WIMP detector technology and exposure will be required in order to observe monthly modulation.  Even in the absence of a detection, however, characterizing the expected signal under the simplest assumptions is still a useful exercise. The annual, diurnal, and monthly modulation describe the most important contributions to the time dependence of the expected dark matter event rate which vary on human time scales.  A full understanding of the predicted event rate improves our ability to discriminate signal from background, for example, by providing null tests for experiments without sufficient exposure to detect monthly modulation.

Nevertheless, given that monthly modulation in WIMP direct-detection experiments is observable in principle and can be used to distinguish dark matter events from background, we considered it in some detail in this work. We examined both the motion of the Earth around the Earth-Moon barycenter and the gravitational focusing due to the Moon and found that the former was the dominant contribution to monthly modulation, being almost completely unaffected by the latter. In addition to the expected monthly cycles in the rate count, the annual envelope is a unique marker that would aid in characterizing a detected signal. Though the expected amplitude of monthly modulation is quite small and thus difficult to detect, any detection would provide distinct, model-independent support for an interpretation of a modulating event rate in terms of dark matter.

On the other hand, monthly modulation of the signal frequency in WISP dark matter searches is large enough to be detectable with current technology and in general cannot be neglected in high resolution searches like those currently underway with ADMX.  Both the motion of the Earth around the Earth-Moon barycenter and the time-dependence of the gravitational focusing due to the eccentricity of the Moon's orbit contribute to the monthly modulation of the frequency.  While these effects are less important for lower resolution scans or for broadband WISP searches, they will become very important for a detailed study of the dark matter velocity distribution in the event of a detection of WISP dark matter.

\section*{Acknowledgments}
We would like to thank Daniel Green and James Owen for helpful discussions.  This research was supported by the Natural Sciences and Engineering Research Council (NSERC) of Canada.

\appendix
\section{Coordinates}\label{coordinates}

\begin{figure}[h!]%
\centering
\includegraphics[width=0.8\textwidth]{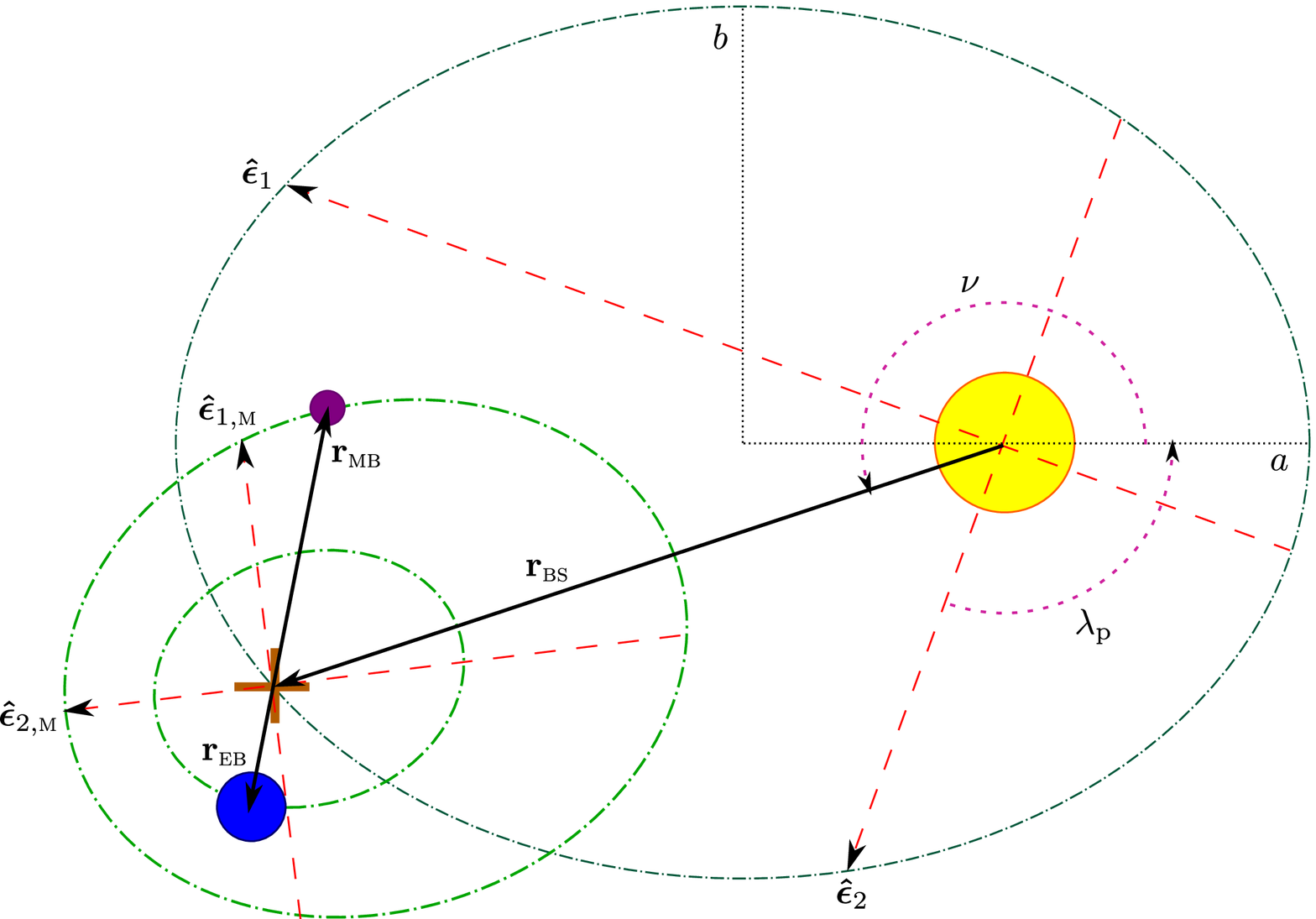}%
\caption{Schematic drawing of orbits of Earth, Moon, and Earth-Moon barycenter.  The Earth-Moon barycenter, represented by a brown cross, orbits the Sun (yellow) in an ellipse which is spanned by $\boldsymbol{\hat{\epsilon}}_{1}$ and $\boldsymbol{\hat{\epsilon}}_{2}$.  Meanwhile, the Earth (blue) and Moon (violet) orbit the Earth-Moon barycenter on similar ellipses which lie in a plane spanned by $\boldsymbol{\hat{\epsilon}}_{1,\textsc{m}}$ and $\boldsymbol{\hat{\epsilon}}_{2,\textsc{m}}$.  This diagram is not to scale.}%
\label{coordinatesfig}%
\end{figure}
In this appendix, we describe the various position vectors, velocity vectors, and coordinate systems that appear throughout this work.  We follow the treatment in \cite{Lee:2013xxa}, adopting the same conventions and notation.  All times are measured in days relative to J2000.0.  We describe the orbits in terms of mean orbital elements which we will approximate as constants.  This is a reasonable approximation for the description of the orbit of the Earth-Moon barycenter about the Sun, but is a poor approximation for the description of the orbits of the Earth and Moon about the Earth-Moon barycenter.  This will not affect the main points presented in the paper, but more accurate results could be obtained by including the evolution of the orbital elements or relying instead on ephemeris data for the positions and velocities of the Earth and Moon.

For a classical Keplerian orbit, given the length of the semi-major axis $a$, and the eccentricity $e$, the central body is located at a distance $f=ae$ from the center of the ellipse. If $t_\T{p}$ denotes the time of periapsis (closest approach) of the orbiting body, the mean anomaly $g(t)$ evolves via
\begin{equation}
	g(t)=\omega (t - t_\T{p})\ ,
\label{g(t)}
\end{equation}
where $\omega=2\pi/T$ and $T$ is the orbital period.  The true anomaly $\nu$, which is the geometric angle in the plane of the ellipse between the periapsis and the orbital body at time $t$, is given by
\begin{equation}
	\nu \simeq g(t)+2e\sin{g(t)}+\frac{5}{4}e^2 \sin{2g(t)}\ ,
\label{nu}
\end{equation}
where we have kept terms through order $e^2$. The distance between the bodies at a given $\nu$ is specified by the relation
\begin{equation}
	r(t)=\frac{a(1-e^2)}{1+e\cos{\nu}}\ ,
\label{dist(t)}
\end{equation}
and the longitude $\lambda(t)$ is
\begin{equation}
	\lambda(t)=\lambda_\T{p} + \nu\ ,
\label{lambda(t)}
\end{equation}
where $\lambda_\T{p}$ is the longitude of the periapsis. The position of orbiting body then, is given by
\begin{equation}
	\mathbf{r}(t)= r(t)\left(-\sin {\lambda(t)}\, \hat{\textbf{\i}} + \cos {\lambda(t)}\, \hat{\textbf{\j}}\right) \ ,
\label{rES}
\end{equation}
where $\hat{\textbf{\i}}$ and $\hat{\textbf{\j}}$ are orthonormal unit vectors that span the plane of the orbit, and $\hat{\textbf{\j}}$ is taken to be the reference direction.

We now describe the motion of the Earth relative to the Sun in two parts: first we will specify the motion of the Earth-Moon barycenter about the Sun, then we will detail the motion of the Earth (and Moon) about the Earth-Moon barycenter.  For the purposes of this calculation, we will assume that the barycenter of the Solar System remains fixed at the center of the Sun.

Using the equations from above, the motion of the Earth-Moon barycenter relative to the Sun is given by
\begin{equation}
	\mathbf{r}_{\textsc{bs}}(t)= r_{\textsc{bs}}(t)\left(-\sin {\lambda_{\textsc{bs}} (t)}\, \boldsymbol{\hat{\epsilon}}_{1} + \cos {\lambda_{\textsc{bs}} (t)}\, \boldsymbol{\hat{\epsilon}}_{2}\right) \ ,
\label{rBS}
\end{equation}
where $\boldsymbol{\hat{\epsilon}}_{1}=(0.9940,0.1085,0.003116)$ and $\boldsymbol{\hat{\epsilon}}_{2}=(-0.05173,0.4945,-0.8677)$ are orthonormal unit vectors (given in Galactic coordinates) that span the ecliptic plane, and the relevant orbital elements are $a_\textsc{bs}=1.4960\times 10^8$~km, $e_\textsc{bs}=0.016722$, $T_\T{yr}= 365.256$~days, $t_{\T{p},\textsc{bs}}=1.70833$~days, and $\lambda_{\T{p},\textsc{bs}}=102.937^\circ$ \cite{Lee:2013xxa,McCabe:2013kea}.

There are a few additional complications in describing the motion of the Earth about the Earth-Moon barycenter.  First, Eq.~(\ref{dist(t)}) gives the distance between the orbiting bodies, but we would instead here like to know the position of the Earth relative to the barycenter, not the Moon.  This distance is in fact given by
\begin{equation}
	r_\textsc{eb}(t)=\frac{r_{\textsc{em}}(t)}{1+\frac{M_\textsc{e}}{M_\textsc{m}}} \ ,
	\label{reb(t)}
\end{equation}
and a similar relation describes the distance between the Moon and the barycenter:
\begin{equation}
	r_\textsc{mb}(t)=\frac{r_{\textsc{em}}(t)}{1+\frac{M_\textsc{m}}{M_\textsc{e}}} \ .
	\label{rem(t)}
\end{equation}
Next, the Earth and Moon orbit about the Earth-Moon barycenter on similar ellipses which lie in a plane which is inclined relative to the ecliptic.  To define their orbits then, we will need to define new unit vectors, $\boldsymbol{\hat{\epsilon}}_{1,\textsc{m}}$ and $\boldsymbol{\hat{\epsilon}}_{2,\textsc{m}}$, which span their barycentric orbital plane. To construct these unit vectors, we begin with $\boldsymbol{\hat{\epsilon}}_1$ and $\boldsymbol{\hat{\epsilon}}_2$ and perform two rotations. The first is a clockwise rotation by the longitude of ascending node $\Omega_\textsc{em}=125.08^\circ$ of the Moon's orbit with respect to the ecliptic, about the vector $\boldsymbol{\hat{\epsilon}}_{3} \equiv \boldsymbol{\hat{\epsilon}}_{1} \times \boldsymbol{\hat{\epsilon}}_{2}$. Then, if we denote as $\boldsymbol{\hat{\epsilon}}_{2}^\prime$ the vector resulting from the rotation of $\boldsymbol{\hat{\epsilon}}_{2}$, the second rotation is anti-clockwise by the angle of inclination $i_\textsc{em}=5.16^\circ$ about $\boldsymbol{\hat{\epsilon}}_{2}^\prime$ \cite{Standish:2001iom}. In terms of rotation matrices,
\begin{equation}
\left(\begin{array}{c}
	\boldsymbol{\hat{\epsilon}}_{1,\textsc{m}} \\ \boldsymbol{\hat{\epsilon}}_{2,\textsc{m}} \\ \boldsymbol{\hat{\epsilon}}_{3,\textsc{m}}
\end{array}\right) = 
\begin{pmatrix}
	\cos{i_\textsc{em}} & 0 & -\sin{i_\textsc{em}} \\ 0 & 1 & 0 \\ \sin{i_\textsc{em}} & 0 & \cos{i_\textsc{em}} 
\end{pmatrix}
\begin{pmatrix}
	\cos{\Omega_\textsc{em}} & \sin{\Omega_\textsc{em}} & 0 \\ -\sin{\Omega_\textsc{em}} & \cos{\Omega_\textsc{em}} & 0 \\ 0 & 0 & 1
\end{pmatrix}
\left(\begin{array}{c}
	\boldsymbol{\hat{\epsilon}}_{1} \\ \boldsymbol{\hat{\epsilon}}_{2} \\ \boldsymbol{\hat{\epsilon}}_{3}
\end{array}\right) \, ,
\label{rotation}
\end{equation}
and so we find $\boldsymbol{\hat{\epsilon}}_{1,\textsc{m}}=(-0.6025,0.2628,-0.7536)$ and $\boldsymbol{\hat{\epsilon}}_{2,\textsc{m}}=(-0.7837,-0.3738,0.4961)$, in Galactic coordinates.

We can now write the position of the Moon relative to the Earth-Moon barycenter in the notation above:
\begin{equation}
	\mathbf{r}_{\textsc{mb}}(t)= r_{\textsc{mb}}(t)\left(-\sin {\lambda_{\textsc{em}} (t)}\,\boldsymbol{\hat{\epsilon}}_{1,\textsc{m}} + \cos {\lambda_{\textsc{em}} (t)}\,\boldsymbol{\hat{\epsilon}}_{2,\textsc{m}}\right) \ ,
\label{rMB}
\end{equation}
where the orbital period is $T_{\mathrm{sid}}= 27.3216$~days, and the orbital elements are $a_\textsc{em}=3.8470\times 10^5$~km, $e_\textsc{em}=0.0554$, $t_{\T{p},\textsc{em}}=18.4493$~days, and $\lambda_{\T{p},\textsc{em}}=318.15^\circ$ \cite{Standish:2001iom}. The position of the Earth relative to the Earth-Moon barycenter is then constructed from the same orbital elements:
\begin{equation}
	\mathbf{r}_{\textsc{eb}}(t)= -r_{\textsc{eb}}(t)\left(-\sin {\lambda_{\textsc{em}} (t)}\,\boldsymbol{\hat{\epsilon}}_{1,\textsc{m}} + \cos {\lambda_{\textsc{em}} (t)}\,\boldsymbol{\hat{\epsilon}}_{2,\textsc{m}}\right) \ .
\label{rEB}
\end{equation}

Given Eqs. (\ref{rBS}) and (\ref{rEB}) then, the description of the Earth's position in the Solar frame, including its motion around the barycenter, is given by
\begin{equation}
	\mathbf{r}_{\textsc{es}}(t)=\mathbf{r}_{\textsc{eb}}(t)+\mathbf{r}_{\textsc{bs}}(t) \ ,
\label{rESB}
\end{equation}
and $\mathbf{v}_{\textsc{es}}(t)=\dot{\mathbf{r}}_{\textsc{es}}(t)$.  Analogously, the position and velocity of the Moon in the Sun's frame are given by
\begin{equation}
	\mathbf{r}_{\textsc{ms}}(t)=\mathbf{r}_{\textsc{mb}}(t)+\mathbf{r}_{\textsc{bs}}(t) \ ;\qquad \mathbf{v}_{\textsc{ms}}(t)=\dot{\mathbf{r}}_{\textsc{ms}}(t) \ .
\label{rMSB}
\end{equation}

\bibliographystyle{unsrtnat}
\bibliography{modulation}

\begin{thebibliography}{57}
\providecommand{\natexlab}[1]{#1}
\providecommand{\url}[1]{\texttt{#1}}
\expandafter\ifx\csname urlstyle\endcsname\relax
  \providecommand{\doi}[1]{doi: #1}\else
  \providecommand{\doi}{doi: \begingroup \urlstyle{rm}\Url}\fi

\bibitem[Olive et~al.(2014)]{Agashe:2014kda}
K.~A. Olive et~al.
\newblock {Review of Particle Physics}.
\newblock \emph{Chin. Phys.}, C38:\penalty0 090001, 2014.
\newblock \doi{10.1088/1674-1137/38/9/090001}.

\bibitem[Jungman et~al.(1996)Jungman, Kamionkowski, and Griest]{Jungman:1995df}
Gerard Jungman, Marc Kamionkowski, and Kim Griest.
\newblock {Supersymmetric dark matter}.
\newblock \emph{Phys.Rept.}, 267:\penalty0 195--373, 1996.
\newblock \doi{10.1016/0370-1573(95)00058-5}.

\bibitem[Bertone et~al.(2005)Bertone, Hooper, and Silk]{Bertone:2004pz}
Gianfranco Bertone, Dan Hooper, and Joseph Silk.
\newblock {Particle dark matter: Evidence, candidates and constraints}.
\newblock \emph{Phys.Rept.}, 405:\penalty0 279--390, 2005.
\newblock \doi{10.1016/j.physrep.2004.08.031}.

\bibitem[Arias et~al.(2012)Arias, Cadamuro, Goodsell, Jaeckel, Redondo, and
  Ringwald]{Arias:2012az}
Paola Arias, Davide Cadamuro, Mark Goodsell, Joerg Jaeckel, Javier Redondo, and
  Andreas Ringwald.
\newblock {WISPy Cold Dark Matter}.
\newblock \emph{JCAP}, 1206:\penalty0 013, 2012.
\newblock \doi{10.1088/1475-7516/2012/06/013}.

\bibitem[Ringwald(2012)]{Ringwald:2012hr}
Andreas Ringwald.
\newblock {Exploring the Role of Axions and Other WISPs in the Dark Universe}.
\newblock \emph{Phys. Dark Univ.}, 1:\penalty0 116--135, 2012.
\newblock \doi{10.1016/j.dark.2012.10.008}.

\bibitem[Essig et~al.(2013)]{Essig:2013lka}
Rouven Essig et~al.
\newblock {Working Group Report: New Light Weakly Coupled Particles}.
\newblock In \emph{{Community Summer Study 2013: Snowmass on the Mississippi
  (CSS2013) Minneapolis, MN, USA, July 29-August 6, 2013}}, 2013.
\newblock URL
  \url{http://inspirehep.net/record/1263039/files/arXiv:1311.0029.pdf}.

\bibitem[Goodman and Witten(1985)]{Goodman:1984dc}
Mark~W. Goodman and Edward Witten.
\newblock {Detectability of Certain Dark Matter Candidates}.
\newblock \emph{Phys.Rev.}, D31:\penalty0 3059, 1985.
\newblock \doi{10.1103/PhysRevD.31.3059}.

\bibitem[Smith and Lewin(1990)]{Smith:1988kw}
P.F. Smith and J.D. Lewin.
\newblock {Dark Matter Detection}.
\newblock \emph{Phys.Rept.}, 187:\penalty0 203, 1990.
\newblock \doi{10.1016/0370-1573(90)90081-C}.

\bibitem[Lewin and Smith(1996)]{Lewin:1995rx}
J.D. Lewin and P.F. Smith.
\newblock {Review of mathematics, numerical factors, and corrections for dark
  matter experiments based on elastic nuclear recoil}.
\newblock \emph{Astropart.Phys.}, 6:\penalty0 87--112, 1996.
\newblock \doi{10.1016/S0927-6505(96)00047-3}.

\bibitem[Asztalos et~al.(2001)]{Asztalos:2001tf}
Stephen~J. Asztalos et~al.
\newblock {Large scale microwave cavity search for dark matter axions}.
\newblock \emph{Phys. Rev.}, D64:\penalty0 092003, 2001.
\newblock \doi{10.1103/PhysRevD.64.092003}.

\bibitem[Asztalos et~al.(2010)]{Asztalos:2009yp}
S.J. Asztalos et~al.
\newblock {A SQUID-based microwave cavity search for dark-matter axions}.
\newblock \emph{Phys.Rev.Lett.}, 104:\penalty0 041301, 2010.
\newblock \doi{10.1103/PhysRevLett.104.041301}.

\bibitem[Slocum et~al.(2015)Slocum, Baker, Hirshfield, Jiang, Malagon, Martin,
  Shchelkunov, and Szymkowiak]{Slocum:2014gwa}
P.~L. Slocum, O.~K. Baker, J.~L. Hirshfield, Y.~Jiang, A.~T. Malagon, A.~J.
  Martin, S.~Shchelkunov, and A.~Szymkowiak.
\newblock {Design and Calibration of the 34 GHz Yale Microwave Cavity
  Experiment}.
\newblock \emph{Nucl. Instrum. Meth.}, A770:\penalty0 76--86, 2015.
\newblock \doi{10.1016/j.nima.2014.10.013}.

\bibitem[Horns et~al.(2014)Horns, Lindner, Lobanov, and
  Ringwald]{Horns:2014qta}
Dieter Horns, Axel Lindner, Andrei Lobanov, and Andreas Ringwald.
\newblock {WISP Dark Matter eXperiment and Prospects for Broadband Dark Matter
  Searches in the $1\,\mu$eV--$10\,$meV Mass Range}.
\newblock In \emph{{10th Patras Workshop on Axions, WIMPs and WISPs (AXION-WIMP
  2014) Geneva, Switzerland, June 29-July 4, 2014}}, 2014.
\newblock URL
  \url{http://inspirehep.net/record/1323442/files/arXiv:1410.6302.pdf}.

\bibitem[Drukier et~al.(1986)Drukier, Freese, and Spergel]{Drukier:1986tm}
A.K. Drukier, K.~Freese, and D.N. Spergel.
\newblock {Detecting Cold Dark Matter Candidates}.
\newblock \emph{Phys.Rev.}, D33:\penalty0 3495--3508, 1986.
\newblock \doi{10.1103/PhysRevD.33.3495}.

\bibitem[Freese et~al.(1988)Freese, Frieman, and Gould]{Freese:1987wu}
Katherine Freese, Joshua~A. Frieman, and Andrew Gould.
\newblock {Signal Modulation in Cold Dark Matter Detection}.
\newblock \emph{Phys.Rev.}, D37:\penalty0 3388--3405, 1988.
\newblock \doi{10.1103/PhysRevD.37.3388}.

\bibitem[Freese et~al.(2013)Freese, Lisanti, and Savage]{Freese:2012xd}
Katherine Freese, Mariangela Lisanti, and Christopher Savage.
\newblock {Annual Modulation of Dark Matter: A Review}.
\newblock \emph{Rev.Mod.Phys.}, 85:\penalty0 1561--1581, 2013.
\newblock \doi{10.1103/RevModPhys.85.1561}.

\bibitem[Bernabei et~al.(2013)Bernabei, Belli, Cappella, Caracciolo,
  Castellano, et~al.]{Bernabei:2013xsa}
R.~Bernabei, P.~Belli, F.~Cappella, V.~Caracciolo, S.~Castellano, et~al.
\newblock {Final model independent result of DAMA/LIBRA-phase1}.
\newblock \emph{Eur.Phys.J.}, C73:\penalty0 2648, 2013.
\newblock \doi{10.1140/epjc/s10052-013-2648-7}.

\bibitem[Akerib et~al.(2014)]{Akerib:2013tjd}
D.S. Akerib et~al.
\newblock {First results from the LUX dark matter experiment at the Sanford
  Underground Research Facility}.
\newblock \emph{Phys.Rev.Lett.}, 112:\penalty0 091303, 2014.
\newblock \doi{10.1103/PhysRevLett.112.091303}.

\bibitem[Aprile et~al.(2012)]{Aprile:2012nq}
E.~Aprile et~al.
\newblock {Dark Matter Results from 225 Live Days of XENON100 Data}.
\newblock \emph{Phys.Rev.Lett.}, 109:\penalty0 181301, 2012.
\newblock \doi{10.1103/PhysRevLett.109.181301}.

\bibitem[Angle et~al.(2011)]{Angle:2011th}
J.~Angle et~al.
\newblock {A search for light dark matter in XENON10 data}.
\newblock \emph{Phys.Rev.Lett.}, 107:\penalty0 051301, 2011.
\newblock \doi{10.1103/PhysRevLett.110.249901, 10.1103/PhysRevLett.107.051301}.

\bibitem[Agnese et~al.(2013)]{Agnese:2013rvf}
R.~Agnese et~al.
\newblock {Silicon Detector Dark Matter Results from the Final Exposure of CDMS
  II}.
\newblock \emph{Phys.Rev.Lett.}, 111:\penalty0 251301, 2013.
\newblock \doi{10.1103/PhysRevLett.111.251301}.

\bibitem[Agnese et~al.(2014)]{Agnese:2013jaa}
R.~Agnese et~al.
\newblock {Search for Low-Mass Weakly Interacting Massive Particles Using
  Voltage-Assisted Calorimetric Ionization Detection in the SuperCDMS
  Experiment}.
\newblock \emph{Phys.Rev.Lett.}, 112\penalty0 (4):\penalty0 041302, 2014.
\newblock \doi{10.1103/PhysRevLett.112.041302}.

\bibitem[Armengaud et~al.(2011)]{Armengaud:2011cy}
E.~Armengaud et~al.
\newblock {Final results of the EDELWEISS-II WIMP search using a 4-kg array of
  cryogenic germanium detectors with interleaved electrodes}.
\newblock \emph{Phys.Lett.}, B702:\penalty0 329--335, 2011.
\newblock \doi{10.1016/j.physletb.2011.07.034}.

\bibitem[Savage et~al.(2009)Savage, Gelmini, Gondolo, and
  Freese]{Savage:2008er}
C.~Savage, G.~Gelmini, P.~Gondolo, and K.~Freese.
\newblock {Compatibility of DAMA/LIBRA dark matter detection with other
  searches}.
\newblock \emph{JCAP}, 0904:\penalty0 010, 2009.
\newblock \doi{10.1088/1475-7516/2009/04/010}.

\bibitem[Ralston(2010)]{Ralston:2010bd}
John~P. Ralston.
\newblock {One Model Explains DAMA/LIBRA, CoGENT, CDMS, and XENON}.
\newblock 2010.

\bibitem[Blum(2011)]{Blum:2011jf}
Kfir Blum.
\newblock {DAMA vs. the annually modulated muon background}.
\newblock 2011.

\bibitem[Davis(2014)]{Davis:2014cja}
Jonathan~H. Davis.
\newblock {Fitting the annual modulation in DAMA with neutrons from muons and
  neutrinos}.
\newblock \emph{Phys.Rev.Lett.}, 113:\penalty0 081302, 2014.
\newblock \doi{10.1103/PhysRevLett.113.081302}.

\bibitem[Bernabei et~al.(2012)Bernabei, Belli, Cappella, Caracciolo, Cerulli,
  et~al.]{Bernabei:2012wp}
R.~Bernabei, P.~Belli, F.~Cappella, V.~Caracciolo, R.~Cerulli, et~al.
\newblock {No role for muons in the DAMA annual modulation results}.
\newblock \emph{Eur.Phys.J.}, C72:\penalty0 2064, 2012.
\newblock \doi{10.1140/epjc/s10052-012-2064-4}.

\bibitem[Bernabei et~al.(2014)Bernabei, Belli, Cappella, Caracciolo, Cerulli,
  et~al.]{Bernabei:2014tqa}
R.~Bernabei, P.~Belli, F.~Cappella, V.~Caracciolo, R.~Cerulli, et~al.
\newblock {No role for neutrons, muons and solar neutrinos in the DAMA annual
  modulation results}.
\newblock 2014.

\bibitem[Barbeau et~al.(2014)Barbeau, Collar, Efremenko, and
  Scholberg]{Barbeau:2014mla}
P.S. Barbeau, J.I. Collar, Yu. Efremenko, and K.~Scholberg.
\newblock {Comment on "Fitting the annual modulation in DAMA with neutrons from
  muons and neutrinos''}.
\newblock 2014.

\bibitem[Collar and Avignone(1992)]{Collar:1992qc}
J.I. Collar and F.T. Avignone.
\newblock {Diurnal modulation effects in cold dark matter experiments}.
\newblock \emph{Phys.Lett.}, B275:\penalty0 181--185, 1992.
\newblock \doi{10.1016/0370-2693(92)90873-3}.

\bibitem[Ling et~al.(2004)Ling, Sikivie, and Wick]{Ling:2004aj}
Fu-Sin Ling, Pierre Sikivie, and Stuart Wick.
\newblock {Diurnal and annual modulation of cold dark matter signals}.
\newblock \emph{Phys.Rev.}, D70:\penalty0 123503, 2004.
\newblock \doi{10.1103/PhysRevD.70.123503}.

\bibitem[Lee et~al.(2013)Lee, Lisanti, and Safdi]{Lee:2013xxa}
Samuel~K. Lee, Mariangela Lisanti, and Benjamin~R. Safdi.
\newblock {Dark-Matter Harmonics Beyond Annual Modulation}.
\newblock \emph{JCAP}, 1311:\penalty0 033, 2013.
\newblock \doi{10.1088/1475-7516/2013/11/033}.

\bibitem[Kouvaris and Shoemaker(2014)]{Kouvaris:2014lpa}
Chris Kouvaris and Ian~M. Shoemaker.
\newblock {Daily Modulation as a Smoking Gun of Dark Matter with Significant
  Stopping}.
\newblock 2014.

\bibitem[Schou et~al.(1998)]{1998ApJ...505..390S}
J.~Schou et~al.
\newblock {Helioseismic Studies of Differential Rotation in the Solar Envelope
  by the Solar Oscillations Investigation Using the Michelson Doppler Imager}.
\newblock \emph{Astrophys. J.}, 505:\penalty0 390--417, September 1998.
\newblock \doi{10.1086/306146}.

\bibitem[Sikivie(1983)]{Sikivie:1983ip}
P.~Sikivie.
\newblock {Experimental Tests of the Invisible Axion}.
\newblock \emph{Phys. Rev. Lett.}, 51:\penalty0 1415--1417, 1983.
\newblock \doi{10.1103/PhysRevLett.51.1415}.
\newblock [Erratum: Phys. Rev. Lett.52,695(1984)].

\bibitem[Horns et~al.(2013)Horns, Jaeckel, Lindner, Lobanov, Redondo, and
  Ringwald]{Horns:2012jf}
Dieter Horns, Joerg Jaeckel, Axel Lindner, Andrei Lobanov, Javier Redondo, and
  Andreas Ringwald.
\newblock {Searching for WISPy Cold Dark Matter with a Dish Antenna}.
\newblock \emph{JCAP}, 1304:\penalty0 016, 2013.
\newblock \doi{10.1088/1475-7516/2013/04/016}.

\bibitem[Kerr and Lynden-Bell(1986)]{Kerr:1986hz}
F.~J. Kerr and Donald Lynden-Bell.
\newblock {Review of galactic constants}.
\newblock \emph{Mon.Not.Roy.Astron.Soc.}, 221:\penalty0 1023, 1986.

\bibitem[Schoenrich et~al.(2009)Schoenrich, Binney, and
  Dehnen]{Schoenrich:2009bx}
R.~Schoenrich, J.~Binney, and W.~Dehnen.
\newblock {Local Kinematics and the Local Standard of Rest}.
\newblock 2009.

\bibitem[Lee et~al.(2014)Lee, Lisanti, Peter, and Safdi]{Lee:2013wza}
Samuel~K. Lee, Mariangela Lisanti, Annika H.~G. Peter, and Benjamin~R. Safdi.
\newblock {Effect of Gravitational Focusing on Annual Modulation in Dark-Matter
  Direct-Detection Experiments}.
\newblock \emph{Phys.Rev.Lett.}, 112\penalty0 (1):\penalty0 011301, 2014.
\newblock \doi{10.1103/PhysRevLett.112.011301}.

\bibitem[Alenazi and Gondolo(2006)]{Alenazi:2006wu}
Moqbil~S. Alenazi and Paolo Gondolo.
\newblock {Phase-space distribution of unbound dark matter near the Sun}.
\newblock \emph{Phys.Rev.}, D74:\penalty0 083518, 2006.
\newblock \doi{10.1103/PhysRevD.74.083518}.

\bibitem[Sikivie and Wick(2002)]{Sikivie:2002bj}
Pierre Sikivie and Stuart Wick.
\newblock {Solar wakes of dark matter flows}.
\newblock \emph{Phys.Rev.}, D66:\penalty0 023504, 2002.
\newblock \doi{10.1103/PhysRevD.66.023504}.

\bibitem[Smith et~al.(2007)Smith, Ruchti, Helmi, Wyse, Fulbright,
  et~al.]{Smith:2006ym}
Martin~C. Smith, G.R. Ruchti, A.~Helmi, R.F.G. Wyse, J.P. Fulbright, et~al.
\newblock {The RAVE Survey: Constraining the Local Galactic Escape Speed}.
\newblock \emph{Mon.Not.Roy.Astron.Soc.}, 379:\penalty0 755--772, 2007.
\newblock \doi{10.1111/j.1365-2966.2007.11964.x}.

\bibitem[Freese et~al.(2005)Freese, Gondolo, and Newberg]{Freese:2003tt}
Katherine Freese, Paolo Gondolo, and Heidi~Jo Newberg.
\newblock {Detectability of weakly interacting massive particles in the
  Sagittarius dwarf tidal stream}.
\newblock \emph{Phys.Rev.}, D71:\penalty0 043516, 2005.
\newblock \doi{10.1103/PhysRevD.71.043516}.

\bibitem[Savage et~al.(2006)Savage, Freese, and Gondolo]{Savage:2006qr}
Christopher Savage, Katherine Freese, and Paolo Gondolo.
\newblock {Annual Modulation of Dark Matter in the Presence of Streams}.
\newblock \emph{Phys.Rev.}, D74:\penalty0 043531, 2006.
\newblock \doi{10.1103/PhysRevD.74.043531}.

\bibitem[Lewis and Freese(2004)]{Lewis:2003bv}
Matthew~J. Lewis and Katherine Freese.
\newblock {The Phase of the annual modulation: Constraining the WIMP mass}.
\newblock \emph{Phys.Rev.}, D70:\penalty0 043501, 2004.
\newblock \doi{10.1103/PhysRevD.70.043501}.

\bibitem[Sikivie(1998)]{Sikivie:1997ng}
P.~Sikivie.
\newblock {Caustic rings of dark matter}.
\newblock \emph{Phys. Lett.}, B432:\penalty0 139--144, 1998.
\newblock \doi{10.1016/S0370-2693(98)00595-4}.

\bibitem[Sikivie(1999)]{Sikivie:1999jv}
Pierre Sikivie.
\newblock {The caustic ring singularity}.
\newblock \emph{Phys. Rev.}, D60:\penalty0 063501, 1999.
\newblock \doi{10.1103/PhysRevD.60.063501}.

\bibitem[Sikivie and Ipser(1992)]{Sikivie:1992bk}
P.~Sikivie and James~R. Ipser.
\newblock {Phase space structure of cold dark matter halos}.
\newblock \emph{Phys. Lett.}, B291:\penalty0 288--292, 1992.
\newblock \doi{10.1016/0370-2693(92)91047-D}.

\bibitem[Natarajan and Sikivie(2005)]{Natarajan:2005fh}
Aravind Natarajan and Pierre Sikivie.
\newblock {Robustness of discrete flows and caustics in cold dark matter
  cosmology}.
\newblock \emph{Phys. Rev.}, D72:\penalty0 083513, 2005.
\newblock \doi{10.1103/PhysRevD.72.083513}.

\bibitem[Duffy et~al.(2005)Duffy, Sikivie, Tanner, Asztalos, Hagmann, Kinion,
  Rosenberg, van Bibber, Yu, and Bradley]{Duffy:2005ab}
Leanne Duffy, P.~Sikivie, D.~B. Tanner, Stephen~J. Asztalos, C.~Hagmann,
  D.~Kinion, L.~J Rosenberg, K.~van Bibber, D.~Yu, and R.~F. Bradley.
\newblock {Results of a search for cold flows of dark matter axions}.
\newblock \emph{Phys. Rev. Lett.}, 95:\penalty0 091304, 2005.
\newblock \doi{10.1103/PhysRevLett.95.091304}.

\bibitem[Duffy et~al.(2006)Duffy, Sikivie, Tanner, Asztalos, Hagmann, Kinion,
  Rosenberg, van Bibber, Yu, and Bradley]{Duffy:2006aa}
Leanne~D. Duffy, P.~Sikivie, D.~B. Tanner, Stephen~J. Asztalos, C.~Hagmann,
  D.~Kinion, L.~J Rosenberg, K.~van Bibber, D.~B. Yu, and R.~F. Bradley.
\newblock {A high resolution search for dark-matter axions}.
\newblock \emph{Phys. Rev.}, D74:\penalty0 012006, 2006.
\newblock \doi{10.1103/PhysRevD.74.012006}.

\bibitem[Hoskins et~al.(2011)]{Hoskins:2011iv}
J.~Hoskins et~al.
\newblock {A search for non-virialized axionic dark matter}.
\newblock \emph{Phys. Rev.}, D84:\penalty0 121302, 2011.
\newblock \doi{10.1103/PhysRevD.84.121302}.

\bibitem[Hasenbalg et~al.(1997)Hasenbalg, Abriola, Avignone, Collar,
  Di~Gregorio, et~al.]{Hasenbalg:1997hs}
F.~Hasenbalg, D.~Abriola, F.T. Avignone, J.I. Collar, D.E. Di~Gregorio, et~al.
\newblock {Cold dark matter identification: Diurnal modulation revisited}.
\newblock \emph{Phys.Rev.}, D55:\penalty0 7350--7355, 1997.
\newblock \doi{10.1103/PhysRevD.55.7350}.

\bibitem[Foot and Vagnozzi(2015)]{Foot:2014osa}
R.~Foot and S.~Vagnozzi.
\newblock {Diurnal modulation signal from dissipative hidden sector dark
  matter}.
\newblock \emph{Phys. Lett.}, B748:\penalty0 61--66, 2015.
\newblock \doi{10.1016/j.physletb.2015.06.063}.

\bibitem[McCabe(2014)]{McCabe:2013kea}
Christopher McCabe.
\newblock {The Earth's velocity for direct detection experiments}.
\newblock \emph{JCAP}, 1402:\penalty0 027, 2014.
\newblock \doi{10.1088/1475-7516/2014/02/027}.

\bibitem[Standish(2001)]{Standish:2001iom}
E.M. Standish.
\newblock {Approximate Mean Ecliptic Elements of the Lunar Orbit}.
\newblock \emph{JPL IOM}, 312.F-01-004, 2001.

\end{thebibliography}
\end{document}